\documentclass[aps, prb, amssymb, amsmath, superscriptaddress, 
twocolumn] {revtex4-2}

\usepackage{chngcntr}

\usepackage{graphicx}
\usepackage{color}
\usepackage{amsmath}
\usepackage{enumitem}
\usepackage{amssymb}
\usepackage[hidelinks]{hyperref}
\usepackage{cancel}
\usepackage{ulem}
\usepackage{multirow}

\newcommand{\be}{\begin{equation}}
	\newcommand{\ee}{\end{equation}}

\newcommand{\bea}{\begin{eqnarray}}
	\newcommand{\eea}{\end{eqnarray}}

\renewcommand{\vec}[1]{{\boldsymbol #1}}

\makeatother

\begin{document}
	\title{Detection of  collective modes in unconventional superconductors using tunneling spectroscopy}
	
\author{Patrick A. Lee}
\affiliation{
Department of Physics, Massachusetts Institute of Technology, Cambridge, MA, USA
}
\author{ Jacob F. Steiner}
\affiliation{
Department of Physics and Institute for Quantum Information and Matter,
California Institute of Technology, Pasadena, CA 91125, USA
}

\begin{abstract}
We propose using tunneling spectroscopy with a superconducting electrode to probe the collective modes of  unconventional superconductors. The modes are predicted to appear as peaks in $dI/dV$ at  voltages given by $eV=\omega_i /2$ where  $\omega_i$ denotes the mode frequencies.  
This may prove to be a powerful tool to investigate the pairing symmetry of unconventional superconductors.  The peaks associated with the collective modes appear at fourth order in the single particle tunneling matrix element.  
At the same fourth order, multiple Andreev reflection (MAR) leads to peaks at voltage equal to the energy gaps, which, in BCS superconductors, coincides with the expected position of the amplitude (Higgs) mode. The peaks stemming from the collective modes of unconventional superconductors do not suffer from this coincidence.  
For scanning tunneling microscopes (STM), we estimate that the magnitude of the collective mode contribution is smaller than the MAR contribution by the ratio of the energy gap to the Fermi energy. Moreover, there is no access to the mode dispersion. Conversely, for planar tunnel junctions the collective mode peak is expected to dominate over the MAR peak, and the mode dispersion can be measured. We discuss systems where the search for such collective modes is promising.

\end{abstract} 

\maketitle

\section{Introduction}

In the past three decades, many examples of unconventional superconductors (SC) have been discovered.  
Many of these have multiple order parameters, either due to pairing in several disconnected Fermi surfaces, or due to pairing that is intrinsically multi-component. In the latter case, the order parameters may be members of a particular irreducible representation, prime examples being MgB$_2$ \cite{blumberg2007observation} and the iron based superconductors \cite{paglione2010high}. Alternatively, they are of mixed symmetry due to the breaking of lattice or time reversal symmetry. While there are numerous examples of mixed symmetry pairing, very often it is difficult to identify the precise order parameter symmetry in these materials.
In an interesting recent paper Poniatowski et al. \cite{poniatowski2022spectroscopic} pointed out that, since these systems exhibit collective modes beyond the familiar phase and amplitude (Higgs) modes, the detection of these modes may serve as signature of the order parameter symmetry. They investigated several examples and showed that commonly these modes lie below the quasi-particle gap $2\Delta$ and hence form well defined excitations. Some of these collective modes are analogs of the Leggett mode \cite{leggett1966number}, or of the ``clapping" mode, familiar from the He$^3$ literature \cite{wolfle1976order,volovik2014higgs}.   
While progress in the detection of such modes has been made using nonlinear optical spectroscopy \cite{katsumi2020superconducting}, they are often charge neutral and thus evade detection using conventional tools.
Motivated by this, we study the question whether the collective modes of unconventional superconductors may be detected using tunneling spectroscopy. We investigate  point contact tunneling such as scanning tunneling microscopy (STM) as well as planar tunneling and compare their respective advantages and disadvantages. We will also discuss examples where such experiments may be feasible. 
 
This paper is structured as follows. In Sec.\,\ref{sec:collective_modes_in_tunneling} we demonstrate within linear response theory how collective modes of the pairing give rise to features in the tunneling current between two SCs. In Sec.\,\ref{sec:josephson_peak_and_mar} we discuss processes that give rise to features in the tunneling current at the same order in perturbation theory which are commonly observed, namely the Josephson peak and multiple Andreev reflections (MAR). In Sec.\,\ref{sec:microscopics} we present a microscopic treatment of the current due to collective modes, which includes the possibility of multi-component order parameters. In Sec.\,\ref{sec:Visibility} we discuss the magnitude of the features in the collective mode relative to the Josephson peak and the leading order MAR contribution. We proceed in Sec.\,\ref{sec:experimental_platforms} by applying our results to several experimentally relevant multi-component SCs, highlighting examples where it is promising to look for the current due to collective modes. Finally, in we conclude Sec.\,\ref{sec:discussion} with a discussion of   the experimental challenges associated with measuring the current due to collective modes.

\begin{figure}
	\centering
	\includegraphics[width=0.4\textwidth]{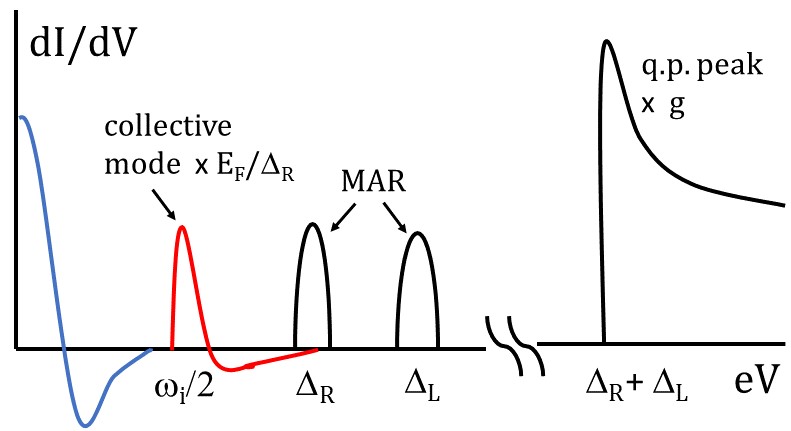}
	\caption{ 
Schematic drawing of the STM tunneling conductance $dI/dV$ with a SC tip (with energy gap $\Delta_L$) showing the expected subgap features up to fourth order in the tunneling amplitude. The standard quasi-particle peak starting at $\Delta_R + \Delta_L$ has been reduced by $g=(h/e^2)/R_N$, the dimensionless normal state conductance. Below this energy we find the multiple Andreev reflection (MAR) peaks  at $\Delta_R$ and $\Delta_L$ which overlap the respective amplitude (Higgs) modes. Shown in  red is
the contribution from a collective mode for an unconventional
SC on the $R$ side at frequency $\omega_i$. It consists of a peak at $eV = \omega_i/2$
and a tail towards higher voltage. Its height has been multiplied by $E_F/\Delta_R$. Shown in blue is the Josephson current that has been broadened by dissipation. The lineshape
is given by Eq.\,\eqref{currentJ}
for $kT_0 \gg E_J$ which is the typical situation \cite{naaman2004subharmonic} and is much narrower in the opposite limit \cite{steinbach2001direct}. 
The collective mode is the new feature discussed in this paper.}
	\label{fig1}
	\vspace{-2mm}
\end{figure}
 
\section{Collective modes in the tunneling spectrum}
\label{sec:collective_modes_in_tunneling}

The idea of using tunneling to detect pair fluctuations goes back to the seminal papers by Ferrell \cite{ferrell1969fluctuations} and Scalapino \cite{scalapino1970pair}. They were interested in pair fluctuations above the critical temperature $T_c$, and pointed out that the pair fluctuations appear in linear response to an external pairing order parameter, just like magnetization fluctuations appear as the linear susceptibility to an external magnetic field. More specifically, they considered a tunnel junction with voltage bias $V$ between two SCs $L$ (left) and $R$ (right) with different $T_{c,j}$ and energy gaps $\Delta_j$, $j\in \{L,R\}$, where the left SC is assumed to have a higher $T_c$. 
The pair tunneling Hamiltonian is obtained by expanding the Josephson energy $E_J$ of a junction with area $A$ to linear order in  $\Delta_R$, which is then replaced by the pair destruction operator $\hat{\Delta}_R(\vec{r}) = |g_0|\ \psi_{R,\downarrow}(\vec{r})\psi_{R,\uparrow}(\vec{r})$, $g_0$ being the BCS coupling. This gives the coupling Hamiltonian
\be \label{pairH}
H_{\mathrm{pair}}
=  \int d\vec{r}\,  C e^{-i2eV t}\ \hat{\Delta}_R(\vec{r}) + \textrm{h.c.},
\ee
which oscillates at the Josephson frequency $2eV$ (setting $\hbar = 1$). Here, we defined the coupling strength $C=\partial (E_J/A)/\partial \Delta_R$  as well as the Josephson energy $E_J=(g/4\pi) \Delta_R K(\sqrt{1-\Delta_R^2/\Delta_L^2})$  in terms of the elliptic function $K$. Moreover, we  introduced the dimensionless conductance $g=(h/e^2)/R_N$, where  $R_N$ is the junction resistance in the normal state \cite{ambegaokar1963tunneling}. In the limit $\Delta_L\gg\Delta_R$, $E_J=(g/2\pi )\Delta_R \ln(4|\Delta_L/\Delta_R|)$ and we recover the expressions given in Ref.\,\cite{scalapino1970pair}. Standard linear response theory gives the current as
\be\label{current1}
I_{\mathrm{pair}}(V,H)=4e C^2 A \ \mathrm{Im}\, \chi_R(\vec{q}=2\vec{q}_H,\omega =2eV),
\ee
where  $\chi_R(\vec{q},\omega)$ is the Fourier transform of the pair susceptibility
\be\label{propagator}
\chi_R (\vec{r},t)= -i\langle [\hat{\Delta}^{}_R(\vec{r},t),\, \hat{\Delta}^{}_R(0)^\dagger] \rangle\theta(t),
\ee
and $2\vec{q}_H$ is the pair momentum induced by a magnetic field $H$  parallel to the junction \cite{scalapino1970pair}. 
(We have assumed that $R$ is a two dimensional SC. Otherwise, the current will depend on the thickness of $R$ provided it is less than the coherence length, c.f. Ref.\,\cite{scalapino1970pair}.) 

The pair fluctuations were successfully measured very close to the transition temperature $T_{c,R}$ \cite{anderson1970experimental}. In principle, there is no reason why the same arguments cannot be applied to low temperatures. In that case, the collective mode dispersions corresponds to poles in $\chi_R(\vec{q},\omega)$ and should manifest as peaks in the tunneling current. In fact, this has been proposed as a way to measure the Higgs mode \cite{pekker2015amplitude}. In practice, 
peaks corresponding to putative collective modes have never been observed in tunneling experiments. A purpose of this paper is to explain this absence, and to point out the conditions under which such observations may become successful in the future.
 
\section{Josephson peak and multiple Andreev reflections}
\label{sec:josephson_peak_and_mar} 

We begin by noting that the current $I_\textrm{pair}$ is proportional to $g^2$ and thus fourth order in the tunneling matrix element. 
We first consider the STM case and  consider other terms to the same order in the tunneling current. 
STM spectra exhibit sub-gap structures Stemming from processes commonly known as multiple Andreev reflections (MAR) \cite{naaman2004subharmonic,ternes2006subgap}.
They may be calculated in an expansion in powers of the tunneling matrix element \cite{cuevas1996hamiltonian}. At fourth order, the first set of MAR peaks appears at $eV=\Delta_L$ and $\Delta_R$ in $dI/dV$. They correspond to processes where a pair tunnels across the junction and gains an energy $2eV$. For $2eV > 2\Delta_R$ this energy can go into exciting a pair of quasi-particles on the $R$ side. 
This gives rise to a step threshold in the current $I(V)$ and consequently a peak in $dI/dV$ at $eV=\Delta_R$. A similar argument produces a step at $\Delta_L$. 
MAR peaks are commonly seen in  STM when the tip is brought close to the surface, increasing $g$ \cite{naaman2004subharmonic,ternes2006subgap}. The ratio of the lowest order MAR peak in $dI/dV$ to the conductance above the coherence peak threshold is simply of order $g$ (see Fig.\,\ref{fig1}).
We note that $g$ of order 0.01 or even unity can be achieved \cite{zhu2020nearly,ternes2006subgap}. Nevertheless, collective modes such as the Higgs mode have not been reported in STM experiments. One reason lies in the fact that, in conventional SC, the phase mode is pushed up to the plasma frequency and the only remaining collective mode is the Higgs mode which has energy 2$\Delta_R$. This gives rise to a peak at $eV=\Delta_R$ which happens to coincide with the lowest order MAR peak. Furthermore, as shown below, in STM the magnitude of the collective mode contribution is reduced from the MAR magnitude by a factor $\Delta_R/E_F$ where $E_F$ is the Fermi energy of the $R$ SC. This reduction stems from point tunneling: in this case, the current involves a convolution over the momentum of the mode. The latter disperses rapidly on the scale of the inverse coherence length $\xi^{-1}$, giving rise to this suppression. We conclude that the collective mode may be visible in STM only for strongly correlated materials where the factor $\Delta_R/E_F$ is not too small. 

Another contribution to the same order in $g^2$ commonly seen in STM is the Josephson current broadened by thermal noise. Thermal fluctuations dephase the junction and convert the Josephson current from a delta function to a peak structure at low but finite bias. The theory has been given by Ivanchenko and Zilberman \cite{ivanchenko1969josephson}. The result depends on the relative size of the Josephson energy $E_J$ of the junction to the noise temperature $k_BT_0$. (Note that $T_0$ is in general different from and larger than the sample temperature $T$.) In the limit $E_J \ll k_BT_0$ the current is given by 
\be\label{currentJ}
I_J(V)= e \frac{E_J^2}{k_B T_0}\frac{2eV\Gamma_0}{(2eV)^2+\Gamma_0^2},
\ee
where the width is given by $\Gamma_0=k_B T_0 R_0 (2e)^2$ with the dissipation parametrized by an effective resistance $R_0$. More specifically, the external circuit is modelled by a series resistance $R_0$ (not to be confused with the normal state junction resistance $R_N$) which gives rise to voltage fluctuations across the junction characterized by $\langle \delta V(t) \delta V(t') \rangle = 2k_B T_0 R_0 \delta(t-t')$. Note that  in Eq.\,\eqref{currentJ} the maximum current is  proportional to $E_J^2/T_0$ which is proportional to $g^2$. In fact, STM data are usually in this limit: the lineshape predicted by Eq.\,\eqref{currentJ} is often seen as a peak in $dI/dV$ whose height is comparable to and scales in the same way as the MAR peak at $eV=\Delta_R $ with changing tip height \cite{naaman2004subharmonic}. On the other hand, planar junctions are in the opposite limit $E_J \gg k_BT_0$ because $E_J$ scales with the area. In this case, the peak is very narrow and steep \cite{steinbach2001direct}. A useful physical picture is that of an overdamped particle moving in a tilted ``washboard potential".  
In the limit $E_J \ll k_BT_0$ thermal fluctuations lead to rapid jumps over the washboard barrier and give rise to phase slips, resulting in Eq.\,\eqref{currentJ}. 

In the case of the Josephson current, the voltage bias is small and the washboard is relatively flat. In the case of the collective mode we are in a large voltage regime where the phase is running rapidly down the washboard and subject to weak modulation due to $E_J$. In this case the phase across the junction is given to a good approximation by $\theta(t) \approx 2e[V_{\textrm{ext}}t + \int_0^t dt’ \delta V(t’)]$. Hence, the fluctuating part of the phase correlation is given by $\langle(\delta\theta(t)-\delta\theta(0))^2\rangle=(2e)^2 2k_BT_0R_0 t$. Inserting this into Eq.\,\eqref{propagator}, we find that the effect of thermal noise is to introduce an additional Lorentzian convolution to the response function, with a width given by $\Gamma_0$. Similar arguments show that the width of the MAR peak is  also given by $\Gamma_0$ (see Appendix B). 
Thus, the minimal width of all the sub-gap structures shown in Fig.\,\ref{fig1} is set by the width of the Josephson peak in Eq.\,\eqref{currentJ} which can be readily measured. It follows that a condition for the visibility of the collective mode is simply that its width given by $\Gamma_0$ is not so large that it will overlap other features such as the Josephson peak or the MAR peak. Note that the width of the collective mode and the MAR peak are the same whether $k_BT_0$ is large or small compared with $E_J$. Only the Josephson peak is affected by this condition.
 
\begin{figure}
	\centering
	\includegraphics[width=0.45\textwidth]{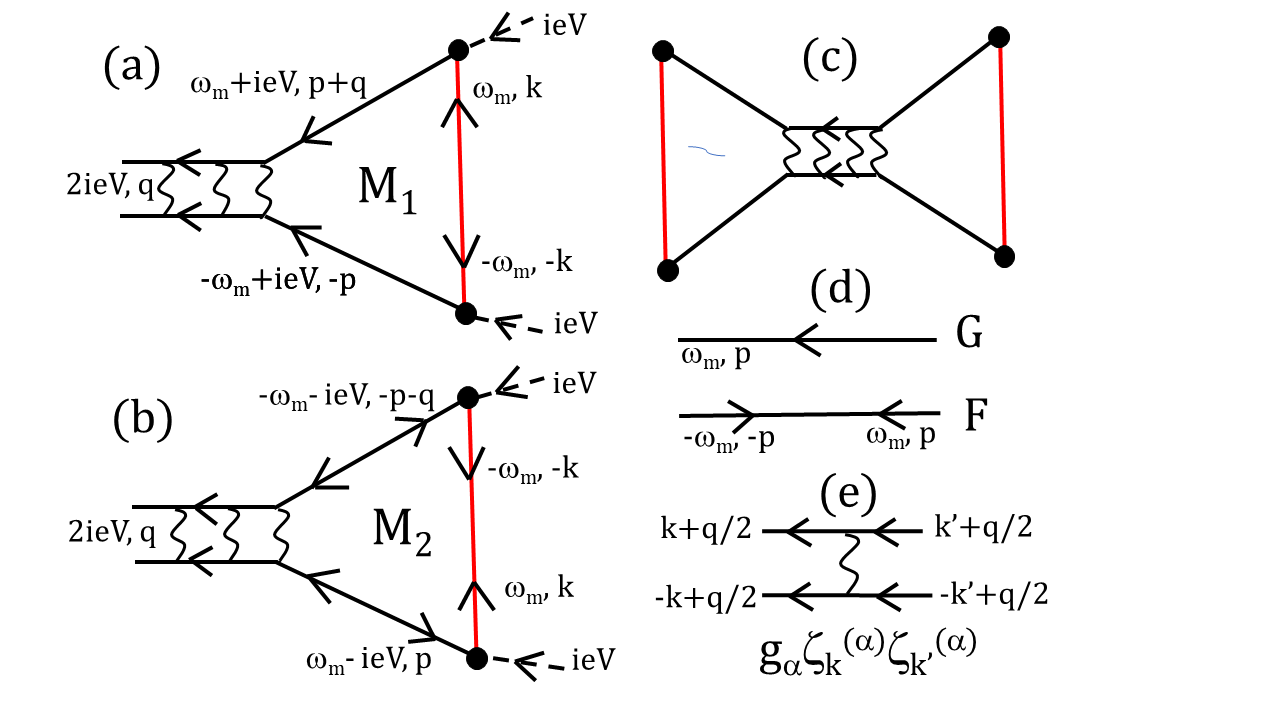}
	\caption{ 
Diagrams that contribute to the STM tunneling current to fourth order in the tunneling matrix  element $t_{\vec{k},\vec{p}}$ represented by the solid dot. The diagrams that couple to the collective mode is shown in (c) where the double line represents the pair propagator of the $R$ SC which is related to the pair susceptibility $\chi_{\alpha,\beta}(\vec{q},\omega = 2eV)$ by analytic continuation. (a) and (b) show the two  diagrams $M_1$ and $M_2$ that contribute to the triangle on the right side of (c). Two similar diagrams contribute to the left triangle related to the right triangle by complex conjugation. The anomalous Green function of the $L$ SC is shown in red. In (a) the $R$ SC Green function is the regular $G$ while in (b) it is the anomalous $F$ function. The latter are shown in (d). Note that frequency and momentum reverse sign on opposite ends of $F$. (e) shows the BCS coupling in separable form for each channel $\alpha$.}
	\label{fig2}
	\vspace{-2mm}
\end{figure}
 
\section{Microscopic treatment of the current due to collective modes}
\label{sec:microscopics} 

Next, we turn to a microscopic treatment of the problem, including the multi-component SCs mentioned in the introduction. 
We will derive an extension of the pair tunneling Hamiltonian Eq.\,\eqref{pairH} by calculating the in-gap current to fourth order in $t_{\vec{k},\vec{p}}$, following earlier work by Takayama \cite{takayama1971superconducting}. The voltage drop across the junction can be absorbed into a time dependent tunneling matrix element $t_{\vec{k},\vec{p}} e^{ieVt}$, rendering the SC leads at equilibrium. This allows a treatment within the conventional Matsubara formalism, and the more elaborate Keldysh treatment \cite{cuevas1996hamiltonian} is not necessary.  The ingap current due to collective modes of the $R$ SC is then given by the diagram Fig.\,\ref{fig2}(c). 
Details of the calculation are presented in Appendix A. Below, we discuss the main results.

In the STM case, tunneling occurs at a single point and does not resolve the momentum of the pair response function as in Eq.\,\eqref{current1}. Instead, the current  involves an integral over the momentum $\vec{q}$:
\begin{multline}\label{currentstm}
I_{\textrm{STM}}(V)=4e
 \sum_{\alpha,\beta}\int d\vec q \ M_\alpha^*(\vec q,V ) M^{}_\beta(\vec q,V )\\ \times \mathrm{Im}\,\chi_{\alpha,\beta}(\vec q,\omega=2eV).
\end{multline}
Here, we have introduced multiple pairing order parameters $\Delta_\vec{k}^{(\alpha)}$ for the $R$ SC. We will drop the $R$ label from now on. The label $\alpha$ may refer to pairing in different bands, or to members of a irreducible representation, or to superposition of different pairing symmetries when time reversal or crystalline symmetry is spontaneously broken. Following Ref.\,\cite{poniatowski2022spectroscopic}, we assume a separable form for the attractive interaction $U_{\vec{k},\vec{k'}}=-\sum_\alpha g_\alpha\zeta_\vec{k}^{(\alpha)} \zeta_\vec{k'}^{(\alpha)}$, where the $\zeta_\vec{k}^{(\alpha)}$ are orthonormal form factors and the $g_\alpha$ are the coupling constants in the corresponding channel. We neglect the dependence on $\vec{q}$, the center of mass momentum of the Cooper pair. This vertex is shown in Fig.\,\ref{fig2}(e).  The pair destruction operator is generalized to $\hat{\Delta}_\vec{k}^{(\alpha)}(\vec{q})=g_\alpha \sum_\vec{k}  \zeta_\vec{k}^{(\alpha)} c^{(\alpha)}_{-\vec{k},\downarrow} c^{(\alpha)}_{\vec{k}+ \vec{q},\uparrow}$.
For simplicity of notation, we assume singlet pairing, but  the calculation can be straightforwardly extended to general pairing symmetry. The pair susceptibility can be generalized from Eq.\,\eqref{propagator} in a natural way as
\be\label{pair2}
\chi_{\alpha,\beta}(\vec{q},t)=-i\langle[\hat{\Delta}_{\vec{k}}^{(\alpha)}
(\vec{q},t) ,  \hat{\Delta}_\vec{k}^{(\beta) }(\vec{q,}0)^\dagger]\rangle\theta(t).
\ee
The matrix element $M_\alpha(\vec q,V )$ is given by the sum of $M_1$ and $M_2$, which are defined by the two triangle diagrams shown in Fig.\,\ref{fig2}(a) and (b), respectively. It is 
\begin{multline}\label{M}
M_\alpha(\vec q,V )=T\sum_{\omega_m } \int d\vec k\, d\vec p \,|\Tilde{t}|^2\zeta_\vec{p}^{(\alpha)}  F_L(\vec k,\omega_m) 
\\
\times [ G_R^{(\alpha)} (\vec {p}+\vec {q}, \omega_m +ieV) G_R^{(\alpha)} (-\vec {p}, -\omega_m +ieV) \\
-F_R^{(\alpha)} (\vec {p}+\vec {q}, \omega_m +ieV) F_R^{(\alpha)} (-\vec {p}, -\omega_m +ieV)].
\end{multline} 
Here, as appropriate for STM tunneling, we have neglected the momentum dependence of the tunneling matrix element $t_{\vec{k},\vec{p}}$, replacing it by $\Tilde{t}$.  
We have further used the fact that the anomalous electron Green function satisfies $F_{\downarrow \uparrow}(\vec k,\omega_m)=-F_{\uparrow \downarrow}
(\vec k,\omega_m)$ 
which accounts for the negative sign in the second term.  
Note that one factor of $\zeta_\vec{k}^{(\alpha)}$ enters the matrix element in Eq.\,\eqref{M} and one factor enters the pair susceptibility in Eq.\,\eqref{currentstm}. It is easy to see that the left triangle in Fig.\,\ref{fig2}(c) is the complex conjugate of the right triangle. The product $\zeta_\vec{p}^{(\alpha)} F_L(\vec{k},\omega_m)$ in Eq.\,\eqref{M} determines which components of the pair fluctuation can be probed. For example, if $L$ is a conventional s-wave SC, only collective modes with a component $\alpha$ corresponding to s-wave will couple, as we shall illustrate by an example below. 

Based on the form Eq.\,\eqref{currentstm}, we suggest a nonlocal generalization of the pair tunneling Hamiltonian,
\be \label{pairH2}
\Tilde{H}_{\mathrm{pair}}
= \sum_\alpha \int d\vec{r}'\,  \Tilde{C}_\alpha(\vec{r}-\vec{r'},V)e^{-i2eVt}\ \hat{\Delta}^{(\alpha)}_R(\vec{r'}) + \textrm{h.c.},
\ee
for STM tunneling at position $\vec{r}$. (Integration over $\vec{r}$ in Eq.\,\eqref{pairH2} gives the generalization of Eq.\,\eqref{pairH} for planar junctions.) It is clear that linear response based on Eq.\,\eqref{pairH2} leads to Eq.\,\eqref{currentstm} if we identify the Fourier transform of $\tilde{C}_\alpha(\vec{r},V)$ with $M_\alpha(\vec{q},V)$. As shown in  Appendix A, $M$ has  a smooth  $V$ dependence which  can usually be ignored. More importantly, Eq.\,\eqref{currentstm} involves a convolution in momentum space between the pair susceptibility and the product of the matrix element. We find  that $M_\alpha(\vec q ,V)$ goes to a constant for small $q$ and falls off with $q$ on a scale given by the inverse of the coherence length $\xi_R$ when  $\Delta_L \gtrsim \Delta_R$ (see Appendix for details). The physical origin of the nonlocality in Eq.\eqref{pairH2} and the convolution over $\vec q$ in Eq.\,\eqref{currentstm} is that the Cooper pair is injected from the $L$ SC one electron at a time by the single particle tunneling matrix element. Consequently quasi-particles exists virtually over a distance of order $\xi_R$  before recombining to form a Cooper pair on the $R$ SC. 
We note that for the single order parameter case the simple pair tunneling Hamiltonian Eq.\,\eqref{pairH} can be readily derived from Eq.\,\eqref{pairH2}. Focusing on the planar case, in the absence of a magnetic field, and for $V \ll \Delta_R$, it is $\tilde{C} \simeq M(\vec{q} = 0,V = 0)/A = \partial (E_J/A) / \partial \Delta_R$ (see Appendix A), fully consistent with Eq.\,\eqref{pairH}. 
 
\section{Visibility }
\label{sec:Visibility} 

Next, we estimate the magnitude of the collective mode contribution to the current. For simplicity we discuss the single order parameter case. In this case, the inverse pair propagator generically takes the form $\chi^{-1} (\vec{q},\omega_n)=N(0)[1 + (\omega_n^2+b_i v_F^2q^2)/(a_i\Delta^2)]$, where
$a_i$ and $b_i$ are numbers of order unity \cite{poniatowski2022spectroscopic}. The zeros of this function give the collective mode dispersion $\omega_i(q)$. The pair susceptibility thus takes the form
\be  \label{Imchi}
\mathrm{Im}\ \chi(\vec{q}, \omega) =\frac{1}{N(0)}\frac{\pi \omega_{i0}^2}{2\omega_i(\vec{q})}\delta(\omega-\omega_i(\vec{q})),
\ee
where $\omega^2_i(\vec{q})=\omega_{i0}^2+b_i v_F^2q^2$ and $\omega_{i0}=\sqrt{a_i} \Delta$. Note that the dispersion is very steep: $\omega_i(\vec{q})$ roughly doubles in value when $q$ is of order the inverse of the coherence length $\xi=v_F/\pi\Delta$. 
In planar junctions this form leads to delta functions in the current $I(V)$ at $2eV=\omega_i(\vec q)$. 
Thus, planar junction tunneling allows access to the dispersion of the mode. This is not the case for  
which requires an additional integration over $\vec q$.  
Instead of a delta function, the $I(V)$ now features a step at $2eV=\omega_{i0}$ followed by a smooth drop off on a scale set by $M(\vec{q},V)$. The step function gives rise to a delta function in $dI/dV$, 
\begin{align}\label{dIdV} 
\frac{dI_{\textrm{STM}} }{dV}=e^2\frac{8\pi^2 a_i \Delta^2}{N(0) b_i  v_F^2} \left\vert M(0,V)\right\vert^2 \delta(2eV-\omega_{i0}),
\end{align}
which is followed by a negative tail towards larger voltages as sketched in Fig.\,\ref{fig1}. To estimate $M$ in Eq.\,\eqref{dIdV}, it suffices to consider its $V=0$ limit. Focusing on a symmetric junction $\Delta_R = \Delta_L = \Delta$, it is $M  = \partial E_J /\partial\Delta = g/8$ \cite{ambegaokar1963tunneling}. 
With this, we may compare the peak in $dI/dV$ due to the collective mode to the step in the MAR current which is given by $eg^2 \Delta$ \cite{cuevas1996hamiltonian}. (See also Appendix B.) Using $N(0)=m/2\pi$ and $E_F=mv_F^2/2$ we find the relative magnitude of the collective mode and MAR steps to be approximately $\Delta/E_F$ as stated earlier. For conventional SCs this ratio is very small which makes the detection of the collective mode infeasible. Nonetheless, there are now examples of strongly correlated SCs where this ratio is not very small. It is worthwhile to look for the collective mode contribution in STM in such systems. 

The situation is more promising for planar junctions. (See Appendix A for details.) It is useful to consider the ratio of the collective mode current $I_{\textrm{planar}}$ to the current in the normal state at $eV=2\Delta_R$ which is given by $I_N=2\Delta/(eR_N)$. For simplicity, we again consider the symmetric case $\Delta_R=\Delta_L=\Delta$. We find the ratio
\be \label{Iplane_ratio}
\frac{I_{\textrm{planar}}}{I_N} = \frac{\pi^3}{32} \frac{a_i g}{Ak_F^2}\frac{\Delta E_F}{\omega_{i0}} \frac{\Gamma}{(2eV-\omega_{i0})^2+\Gamma^2}.
\ee
Here, we have replaced the delta function in Eq.\,\eqref{Imchi} by a Lorentzian with width $\Gamma$ which is given by $\Gamma_0$ plus other sources of broadening such as inhomogeneity. A similar ratio for the planar MAR current is obtained in Appendix B, where it is found to have parametrically the same prefactor up to numerical constants, but with the Lorentzian in Eq.\,\eqref{Iplane_ratio} replaced by $(eV/\Delta) [ (eV)^2-\Delta^2]^{-1/2}$. The latter should also be broadened by voltage noise and inhomogeneity. Since this form is less singular than the Lorentzian, the collective mode contribution should dominate over the MAR peak in planar junctions, in contrast to the situation in STM. More precisely, with the same broadening $\Gamma$, the peak currents due to the collective mode and the MAR have a ratio of $\sqrt{\Delta/\Gamma}$ which is greater than one. 

We now address the size of the signal from the collective mode contribution given by Eq.\,\eqref{Iplane_ratio}. We interpret $Ak_F^2$ as the number of tunneling channels in a planar junction and use the Landauer formula to define the ratio $\mathcal{T}_{\textrm{eff}}=g/(Ak_F^2)$ as the effective tunneling probability per channel. $\mathcal{T}_{\textrm{eff}}$ gives the intrinsic transparency of a tunnel junction and is generally a very small number. Conversely, the peak value of the Lorentzian is $1/\Gamma_i$, and the ratio $E_F/\Gamma_i$ is a very large number. For a typical planar junction, $\mathcal{T}_{\textrm{eff}}$ is so small that the product is still too small to be observable for reasonable $\Gamma_i$. This may be the reason why neither MAR nor collective modes have been observed in planar junctions. However, as we shall see, the numbers are not too far off, and there may be reasons for optimism. To see this we estimate that for the typical oxide tunnel barrier used in Ref.\,\cite{anderson1970experimental}, the transparency is $\mathcal{T}_{\textrm{eff}} \approx 10^{-8}$ (assuming $A \sim 100^2$nm$^2$, $k_F\sim 1$\AA$^{-1}$, and $R_N \sim 2 \Omega$). In this experiment a fluctuating pair tunneling peak with width of about $1 \mu$eV was readily observed. 
We conclude that a collective mode with width of order $1 \mu$eV should be observable in a conventional oxide planar junction.  because the signal in Eq.\,\eqref{Iplane_ratio} is proportional to the ratio $\mathcal{T}_{\textrm{eff}} /\Gamma_i$. The minimal contribution to the width comes from voltage fluctuations and the corresponding $\Gamma_0$ can be made very small \cite{steinbach2001direct}. In practice, in many of the strongly correlated SC of interest, such as cuprates or iron based SC, local inhomogeneity may lead to significant broadening of the collective mode in  planar junctions. Their detection may require higher tunneling transparency $\mathcal{T}_{\textrm{eff}}$ and possibly smaller area junctions. The latter requirement will reduce the current, making the experiment more challenging. 

Apart from the larger signal compared with STM, we  note that the current itself is predicted to show a narrow peak, so that  the $dI/dV$ signal is the derivative of a Lorentzian which has a distinctive lineshape with a large negative part. Furthermore, the planar junction has the advantage that the dispersion of the collective mode may be probed by applying an in-plane magnetic field. These distinctive  features will provide strong evidence that a collective mode is being observed.  
 
\section{Promising experimental platforms}
\label{sec:experimental_platforms} 

Now, we discuss several multi-component examples where the collective modes may be detected in $I(V)$ or $dI/dV$ as sharp peaks. We note that generalization of Eqs.\,\eqref{dIdV} and \eqref{Iplane_ratio} to the case of multiple order parameters is straightforward. Indeed, one may assume that the pair susceptibility takes the form of a weighted sum of delta functions, $\mathrm{Im}\ \chi_{\alpha,\beta}(\vec{q}, \omega) = \sum_{i} w_{\alpha,\beta}^i (\vec{q})\delta(\omega-\omega_i(\vec{q}))$. This translates directly into a sum of peaks at $2eV = \omega_{i0}$ and weighted by $w_{\alpha,\beta}^i$ in the $dI/dV$ [$I(V)$] for STM [planar] tunneling. We expect the magnitude estimates, which were obtained above in the single order parameter case, to apply also in the more general case. 
For simplicity, we assume that the SC being probed is inversion symmetric. If the $L$ SC is conventional, only s-wave Cooper pairs can be injected from the $L$ electrode and only the $\alpha$-components corresponding to s-wave pairing  survive. Consider first multi-band SCs, where s-wave pairing occurs in two different Fermi surfaces $\alpha=1,2$, as, e.g., in MgB$_2$ and in iron based superconductors. In the latter case the two s-wave components are out of phase, referred to as $s_{\pm}$ state. We expect a collective mode (the Leggett mode) corresponding to the out of phase oscillation of the two order parameters $\Delta^{(1)}$ and $\Delta^{(2)}$. This mode will manifest as a pole in $\chi_{1,2}$. 
The Leggett mode has been observed by Raman scattering in MgB$_2$ at a relatively high energy of 9.2 meV which lies between the two doubled energy gaps \cite{cea2016signature}. Hence, this mode is damped, but it is nonetheless interesting to search for this peak within tunneling spectroscopy. We note that MgB$_2$ planar tunneling junctions have been successfully fabricated \cite{shim2007all}. For the Fe based SCs, the situation is not so clear. 
We note that recently an observation of the Leggett mode was reported in single layer NbSe$_2$ in an experiment using a normal STM tip  \cite{wan2022observation}. Here, the Leggett mode is interpreted as giving an excitation above the gap at energy $\Delta +\omega_i$. Interestingly, the experiment found that $\omega_i/2 \approx 0.7 \Delta$ which places the peak well inside the gap and well below the MAR structure. Note that NbSe$_2$ features a small ratio of $\Delta/E_F$ so that  the STM signal of the Leggett mode is expected to be very small, but perhaps a planar junction experiment can be attempted. 

As a second application we consider the case of a time reversal breaking SC, more specifically of the type $s+id$, i. e. an admixture of s- and d-wave pairing. This case is treated in detail in Ref.\,\cite{poniatowski2022spectroscopic} and here we only summarize the salient features. The important point is that the presence of the s wave component allows us to couple to novel collective modes such as clapping modes. We define the s- and d-wave order parameter components as
$(\Delta^{(0)},\Delta^{(2)})=(\eta_0,-i\eta_2)\Delta_0 e^{i\theta}$, where $\theta$ is the overall pair phase,  and $\eta_0$ and $\eta_2$ are  real numbers satisfying
$\eta_0^2+\eta_2^2=1$. It is convenient to 
introduce $\Delta^{\pm}=\Delta^{(0)}/\eta_0 \pm \Delta^{(2)}/(i\eta_2)$. The saddle point solution occurs at $\Delta^+ = \Delta_0$ and $\Delta^- =0$. Expanding around the saddle point we find the coordinates of the collective modes as
\be\label{clap}
\Delta^+(\vec{r},t) = e^{i\theta} [\Delta_0+h(\vec{r},t)],
\ee
\be\label{clap2}
\Delta^-(\vec{r},t) = e^{i\theta} [a(\vec{r},t)+ib(\vec{r},t)].
\ee
Here, $h$ denotes the amplitude or Higgs mode, while $a$ and $b$ denote two new modes which are generalizations of the clapping mode in $p+ip$ SCs. Poniatowski et al. \cite{poniatowski2022spectroscopic} show that these modes lie at approximately $\sqrt{2}\Delta_0$. We will now show that they appear in the s-wave pair fluctuation channel in Eq.\,\eqref{dIdV}. To this end, we expand 
\be\label{clap4}
\Delta^{(0)}(\vec{r},t) = e^{i\theta} [\eta_0 \Delta_0+\Tilde{\Delta}^{(0)}(\vec{r},t)].
\ee
It is easy to see that the fluctuating part is given by
\be\label{clap3}
\Tilde{\Delta}^{(0)}(\vec{r},t) = \eta_0[h(\vec{r},t)+ a(\vec{r},t)+ib(\vec{r},t)].
\ee
Hence, all three modes will appear in the $\alpha=0$ pair fluctuation component in Eq.\,\eqref{dIdV}. In particular, the generalized clapping modes  will show up as peaks in the vicinity of $eV=\Delta_0/\sqrt{2}$, well separated from the MAR peak at $\Delta_0$. This is shown schematically in Fig.\,\ref{fig1}.
In the iron based SC Ba$_{1-x}$K$_x$Fe$_2$As$_2$, a time reversal breaking SC state appears for $x$ between 0.7 and 0.8 and is suspected to be  an $s+id$ SC \cite{grinenko2020superconductivity}. This is an excellent candidate to search for these collective modes.

We consider a third example of triplet pairing where the time reversal breaking state may be of the $p+ip$ or $p+if$ type. UTe$_2$ may be an example \cite{gu2022detection}. This rather complicated structure preserves inversion in the bulk, but it is possible, indeed likely, that the top layer breaks inversion due to some local structural relaxation. In this case the s-wave  pair injected by the $L$ SC is admixed with the p- and f-wave order parameter in the first layer of the $R$ SC, and the matrix element $M_\alpha$ in Eq.\,\eqref{currentstm} is nonzero for non s-wave components. In this way, collective modes may couple to the current. Indeed, a recent STM experiment using a Nb tip found a subgap peak near the expected energy gap for UTe$_2$ and a peak in $dI/dV$ near zero voltage suggestive of a broadened Josephson current \cite{gu2022detection}. The latter observation points to an admixture of s-wave pairing in the top layer, as we need. Unfortunately, the ratio $\Delta/E_F$ may be too small for STM to observe the collective mode in this system. 
 
\section{Discussion}
\label{sec:discussion} 

We end by discussing the feasibility of probing collective modes of unconventional SC using either planar or STM tunnel junctions. 
It is generally considered difficult to make planar tunnel junctions in these systems, 
but with modern fabrication techniques it may be possible to create nanoscale tunnel junctions with  
high transparency that are free of pin-holes. Another approach may involve stacks of van der Waals materials such as transition metal dichalcogenides (TMD). A variety of insulating TMD may be used to create monolayer barriers between SC layers. In cuprates, a single layer of insulating parent state may be used as tunneling barrier \cite{koren2016observation}. 
On the STM front, it has been demonstrated that it is possible to pick up a piece of layered SC with an STM tip, which then serves as the SC electrode \cite{hamidian2016detection}. 
This holds promise for the present application:
e.g., a cuprate SC tip would allow the collective modes of systems with $d$-wave \cite{barlas2013amplitude} or $d+id$ pairing symmetry to be probed. 
As noted above, the detection of collective modes in STM relies on a relatively large ratio of $\Delta/E_F$. Recently, a range of strongly correlated SC have been discovered, which represent promising candidates for the present proposal.  
Notable examples are twisted bilayer and trilayer graphene, where the ratio is so large that the BEC limit may be reached \cite{park2021tunable}. Another example are the iron based topological SC, which have very small Fermi energy \cite{zhu2020nearly}. An example that is not well understood is the superconductivity observed in YPtBi which involves doping of a quadratic touching band with a very small Fermi energy and a short coherence length \cite{kim2018beyond}. 
We conclude that, while they are challenging, tunneling experiments probing the signatures of the collective modes in unconventional SC are within reach. 
\vspace{1cm }
 
\section*{Acknowledgements} 

We thank Shuqiu Wang and Seamus Davis for sharing their data on UTe$_2$ which stimulated this investigation. We thank Nicholas Poniatowski, Leonid Glazman and Iliya Esin for helpful discussions. PL  acknowledges support by DOE (USA) office of Basic Sciences Grant No. DE-FG02-03ER46076. JFS acknowledges support by the Air Force Office of Scientific Research under award number FA9550-22-1-0339.

\bibliography{ref}

\begin{thebibliography}{30}%
\makeatletter
\providecommand \@ifxundefined [1]{%
 \@ifx{#1\undefined}
}%
\providecommand \@ifnum [1]{%
 \ifnum #1\expandafter \@firstoftwo
 \else \expandafter \@secondoftwo
 \fi
}%
\providecommand \@ifx [1]{%
 \ifx #1\expandafter \@firstoftwo
 \else \expandafter \@secondoftwo
 \fi
}%
\providecommand \natexlab [1]{#1}%
\providecommand \enquote  [1]{``#1''}%
\providecommand \bibnamefont  [1]{#1}%
\providecommand \bibfnamefont [1]{#1}%
\providecommand \citenamefont [1]{#1}%
\providecommand \href@noop [0]{\@secondoftwo}%
\providecommand \href [0]{\begingroup \@sanitize@url \@href}%
\providecommand \@href[1]{\@@startlink{#1}\@@href}%
\providecommand \@@href[1]{\endgroup#1\@@endlink}%
\providecommand \@sanitize@url [0]{\catcode `\\12\catcode `\$12\catcode
  `\&12\catcode `\#12\catcode `\^12\catcode `\_12\catcode `\%12\relax}%
\providecommand \@@startlink[1]{}%
\providecommand \@@endlink[0]{}%
\providecommand \url  [0]{\begingroup\@sanitize@url \@url }%
\providecommand \@url [1]{\endgroup\@href {#1}{\urlprefix }}%
\providecommand \urlprefix  [0]{URL }%
\providecommand \Eprint [0]{\href }%
\providecommand \doibase [0]{https://doi.org/}%
\providecommand \selectlanguage [0]{\@gobble}%
\providecommand \bibinfo  [0]{\@secondoftwo}%
\providecommand \bibfield  [0]{\@secondoftwo}%
\providecommand \translation [1]{[#1]}%
\providecommand \BibitemOpen [0]{}%
\providecommand \bibitemStop [0]{}%
\providecommand \bibitemNoStop [0]{.\EOS\space}%
\providecommand \EOS [0]{\spacefactor3000\relax}%
\providecommand \BibitemShut  [1]{\csname bibitem#1\endcsname}%
\let\auto@bib@innerbib\@empty
\bibitem [{\citenamefont {Blumberg}\ \emph {et~al.}(2007)\citenamefont
  {Blumberg}, \citenamefont {Mialitsin}, \citenamefont {Dennis}, \citenamefont
  {Klein}, \citenamefont {Zhigadlo},\ and\ \citenamefont
  {Karpinski}}]{blumberg2007observation}%
  \BibitemOpen
  \bibfield  {author} {\bibinfo {author} {\bibfnamefont {G.}~\bibnamefont
  {Blumberg}}, \bibinfo {author} {\bibfnamefont {A.}~\bibnamefont {Mialitsin}},
  \bibinfo {author} {\bibfnamefont {B.}~\bibnamefont {Dennis}}, \bibinfo
  {author} {\bibfnamefont {M.}~\bibnamefont {Klein}}, \bibinfo {author}
  {\bibfnamefont {N.}~\bibnamefont {Zhigadlo}},\ and\ \bibinfo {author}
  {\bibfnamefont {J.}~\bibnamefont {Karpinski}},\ }\bibfield  {title} {\bibinfo
  {title} {Observation of leggett’s collective mode in a multiband mgb 2
  superconductor},\ }\href@noop {} {\bibfield  {journal} {\bibinfo  {journal}
  {Physical review letters}\ }\textbf {\bibinfo {volume} {99}},\ \bibinfo
  {pages} {227002} (\bibinfo {year} {2007})}\BibitemShut {NoStop}%
\bibitem [{\citenamefont {Paglione}\ and\ \citenamefont
  {Greene}(2010)}]{paglione2010high}%
  \BibitemOpen
  \bibfield  {author} {\bibinfo {author} {\bibfnamefont {J.}~\bibnamefont
  {Paglione}}\ and\ \bibinfo {author} {\bibfnamefont {R.~L.}\ \bibnamefont
  {Greene}},\ }\bibfield  {title} {\bibinfo {title} {High-temperature
  superconductivity in iron-based materials},\ }\href@noop {} {\bibfield
  {journal} {\bibinfo  {journal} {Nature physics}\ }\textbf {\bibinfo {volume}
  {6}},\ \bibinfo {pages} {645} (\bibinfo {year} {2010})}\BibitemShut {NoStop}%
\bibitem [{\citenamefont {Poniatowski}\ \emph {et~al.}(2022)\citenamefont
  {Poniatowski}, \citenamefont {Curtis}, \citenamefont {Yacoby},\ and\
  \citenamefont {Narang}}]{poniatowski2022spectroscopic}%
  \BibitemOpen
  \bibfield  {author} {\bibinfo {author} {\bibfnamefont {N.~R.}\ \bibnamefont
  {Poniatowski}}, \bibinfo {author} {\bibfnamefont {J.~B.}\ \bibnamefont
  {Curtis}}, \bibinfo {author} {\bibfnamefont {A.}~\bibnamefont {Yacoby}},\
  and\ \bibinfo {author} {\bibfnamefont {P.}~\bibnamefont {Narang}},\
  }\bibfield  {title} {\bibinfo {title} {Spectroscopic signatures of
  time-reversal symmetry breaking superconductivity},\ }\href@noop {}
  {\bibfield  {journal} {\bibinfo  {journal} {Communications Physics}\ }\textbf
  {\bibinfo {volume} {5}},\ \bibinfo {pages} {44} (\bibinfo {year}
  {2022})}\BibitemShut {NoStop}%
\bibitem [{\citenamefont {Leggett}(1966)}]{leggett1966number}%
  \BibitemOpen
  \bibfield  {author} {\bibinfo {author} {\bibfnamefont {A.}~\bibnamefont
  {Leggett}},\ }\bibfield  {title} {\bibinfo {title} {Number-phase fluctuations
  in two-band superconductors},\ }\href@noop {} {\bibfield  {journal} {\bibinfo
   {journal} {Progress of Theoretical Physics}\ }\textbf {\bibinfo {volume}
  {36}},\ \bibinfo {pages} {901} (\bibinfo {year} {1966})}\BibitemShut
  {NoStop}%
\bibitem [{\citenamefont {W{\"o}lfle}(1976)}]{wolfle1976order}%
  \BibitemOpen
  \bibfield  {author} {\bibinfo {author} {\bibfnamefont {P.}~\bibnamefont
  {W{\"o}lfle}},\ }\bibfield  {title} {\bibinfo {title} {Order-parameter
  collective modes in he 3-a},\ }\href@noop {} {\bibfield  {journal} {\bibinfo
  {journal} {Physical Review Letters}\ }\textbf {\bibinfo {volume} {37}},\
  \bibinfo {pages} {1279} (\bibinfo {year} {1976})}\BibitemShut {NoStop}%
\bibitem [{\citenamefont {Volovik}\ and\ \citenamefont
  {Zubkov}(2014)}]{volovik2014higgs}%
  \BibitemOpen
  \bibfield  {author} {\bibinfo {author} {\bibfnamefont {G.}~\bibnamefont
  {Volovik}}\ and\ \bibinfo {author} {\bibfnamefont {M.}~\bibnamefont
  {Zubkov}},\ }\bibfield  {title} {\bibinfo {title} {Higgs bosons in particle
  physics and in condensed matter},\ }\href@noop {} {\bibfield  {journal}
  {\bibinfo  {journal} {Journal of Low Temperature Physics}\ }\textbf {\bibinfo
  {volume} {175}},\ \bibinfo {pages} {486} (\bibinfo {year}
  {2014})}\BibitemShut {NoStop}%
\bibitem [{\citenamefont {Katsumi}\ \emph {et~al.}(2020)\citenamefont
  {Katsumi}, \citenamefont {Li}, \citenamefont {Raffy}, \citenamefont
  {Gallais},\ and\ \citenamefont {Shimano}}]{katsumi2020superconducting}%
  \BibitemOpen
  \bibfield  {author} {\bibinfo {author} {\bibfnamefont {K.}~\bibnamefont
  {Katsumi}}, \bibinfo {author} {\bibfnamefont {Z.~Z.}\ \bibnamefont {Li}},
  \bibinfo {author} {\bibfnamefont {H.}~\bibnamefont {Raffy}}, \bibinfo
  {author} {\bibfnamefont {Y.}~\bibnamefont {Gallais}},\ and\ \bibinfo {author}
  {\bibfnamefont {R.}~\bibnamefont {Shimano}},\ }\bibfield  {title} {\bibinfo
  {title} {Superconducting fluctuations probed by the higgs mode in bi 2 sr 2
  ca cu 2 o 8+ x thin films},\ }\href@noop {} {\bibfield  {journal} {\bibinfo
  {journal} {Physical Review B}\ }\textbf {\bibinfo {volume} {102}},\ \bibinfo
  {pages} {054510} (\bibinfo {year} {2020})}\BibitemShut {NoStop}%
\bibitem [{\citenamefont {Naaman}\ and\ \citenamefont
  {Dynes}(2004)}]{naaman2004subharmonic}%
  \BibitemOpen
  \bibfield  {author} {\bibinfo {author} {\bibfnamefont {O.}~\bibnamefont
  {Naaman}}\ and\ \bibinfo {author} {\bibfnamefont {R.}~\bibnamefont {Dynes}},\
  }\bibfield  {title} {\bibinfo {title} {Subharmonic gap structure in
  superconducting scanning tunneling microscope junctions},\ }\href@noop {}
  {\bibfield  {journal} {\bibinfo  {journal} {Solid state communications}\
  }\textbf {\bibinfo {volume} {129}},\ \bibinfo {pages} {299} (\bibinfo {year}
  {2004})}\BibitemShut {NoStop}%
\bibitem [{\citenamefont {Steinbach}\ \emph {et~al.}(2001)\citenamefont
  {Steinbach}, \citenamefont {Joyez}, \citenamefont {Cottet}, \citenamefont
  {Esteve}, \citenamefont {Devoret}, \citenamefont {Huber},\ and\ \citenamefont
  {Martinis}}]{steinbach2001direct}%
  \BibitemOpen
  \bibfield  {author} {\bibinfo {author} {\bibfnamefont {A.}~\bibnamefont
  {Steinbach}}, \bibinfo {author} {\bibfnamefont {P.}~\bibnamefont {Joyez}},
  \bibinfo {author} {\bibfnamefont {A.}~\bibnamefont {Cottet}}, \bibinfo
  {author} {\bibfnamefont {D.}~\bibnamefont {Esteve}}, \bibinfo {author}
  {\bibfnamefont {M.}~\bibnamefont {Devoret}}, \bibinfo {author} {\bibfnamefont
  {M.}~\bibnamefont {Huber}},\ and\ \bibinfo {author} {\bibfnamefont {J.~M.}\
  \bibnamefont {Martinis}},\ }\bibfield  {title} {\bibinfo {title} {Direct
  measurement of the josephson supercurrent in an ultrasmall josephson
  junction},\ }\href@noop {} {\bibfield  {journal} {\bibinfo  {journal}
  {Physical review letters}\ }\textbf {\bibinfo {volume} {87}},\ \bibinfo
  {pages} {137003} (\bibinfo {year} {2001})}\BibitemShut {NoStop}%
\bibitem [{\citenamefont {Ferrell}(1969)}]{ferrell1969fluctuations}%
  \BibitemOpen
  \bibfield  {author} {\bibinfo {author} {\bibfnamefont {R.~A.}\ \bibnamefont
  {Ferrell}},\ }\bibfield  {title} {\bibinfo {title} {Fluctuations and the
  superconducting phase transition: Ii. onset of josephson tunneling and
  paraconductivity of a junction},\ }\href@noop {} {\bibfield  {journal}
  {\bibinfo  {journal} {Journal of Low Temperature Physics}\ }\textbf {\bibinfo
  {volume} {1}},\ \bibinfo {pages} {423} (\bibinfo {year} {1969})}\BibitemShut
  {NoStop}%
\bibitem [{\citenamefont {Scalapino}(1970)}]{scalapino1970pair}%
  \BibitemOpen
  \bibfield  {author} {\bibinfo {author} {\bibfnamefont {D.~J.}\ \bibnamefont
  {Scalapino}},\ }\bibfield  {title} {\bibinfo {title} {Pair tunneling as a
  probe of fluctuations in superconductors},\ }\href@noop {} {\bibfield
  {journal} {\bibinfo  {journal} {Physical Review Letters}\ }\textbf {\bibinfo
  {volume} {24}},\ \bibinfo {pages} {1052} (\bibinfo {year}
  {1970})}\BibitemShut {NoStop}%
\bibitem [{\citenamefont {Ambegaokar}\ and\ \citenamefont
  {Baratoff}(1963)}]{ambegaokar1963tunneling}%
  \BibitemOpen
  \bibfield  {author} {\bibinfo {author} {\bibfnamefont {V.}~\bibnamefont
  {Ambegaokar}}\ and\ \bibinfo {author} {\bibfnamefont {A.}~\bibnamefont
  {Baratoff}},\ }\bibfield  {title} {\bibinfo {title} {Tunneling between
  superconductors},\ }\href@noop {} {\bibfield  {journal} {\bibinfo  {journal}
  {Physical Review Letters}\ }\textbf {\bibinfo {volume} {10}},\ \bibinfo
  {pages} {486} (\bibinfo {year} {1963})}\BibitemShut {NoStop}%
\bibitem [{\citenamefont {Anderson}\ and\ \citenamefont
  {Goldman}(1970)}]{anderson1970experimental}%
  \BibitemOpen
  \bibfield  {author} {\bibinfo {author} {\bibfnamefont {J.}~\bibnamefont
  {Anderson}}\ and\ \bibinfo {author} {\bibfnamefont {A.~M.}\ \bibnamefont
  {Goldman}},\ }\bibfield  {title} {\bibinfo {title} {Experimental
  determination of the pair susceptibility of a superconductor},\ }\href@noop
  {} {\bibfield  {journal} {\bibinfo  {journal} {Physical Review Letters}\
  }\textbf {\bibinfo {volume} {25}},\ \bibinfo {pages} {743} (\bibinfo {year}
  {1970})}\BibitemShut {NoStop}%
\bibitem [{\citenamefont {Pekker}\ and\ \citenamefont
  {Varma}(2015)}]{pekker2015amplitude}%
  \BibitemOpen
  \bibfield  {author} {\bibinfo {author} {\bibfnamefont {D.}~\bibnamefont
  {Pekker}}\ and\ \bibinfo {author} {\bibfnamefont {C.}~\bibnamefont {Varma}},\
  }\bibfield  {title} {\bibinfo {title} {Amplitude/higgs modes in condensed
  matter physics},\ }\href@noop {} {\bibfield  {journal} {\bibinfo  {journal}
  {Annu. Rev. Condens. Matter Phys.}\ }\textbf {\bibinfo {volume} {6}},\
  \bibinfo {pages} {269} (\bibinfo {year} {2015})}\BibitemShut {NoStop}%
\bibitem [{\citenamefont {Ternes}\ \emph {et~al.}(2006)\citenamefont {Ternes},
  \citenamefont {Schneider}, \citenamefont {Cuevas}, \citenamefont {Lutz},
  \citenamefont {Hirjibehedin},\ and\ \citenamefont
  {Heinrich}}]{ternes2006subgap}%
  \BibitemOpen
  \bibfield  {author} {\bibinfo {author} {\bibfnamefont {M.}~\bibnamefont
  {Ternes}}, \bibinfo {author} {\bibfnamefont {W.-D.}\ \bibnamefont
  {Schneider}}, \bibinfo {author} {\bibfnamefont {J.-C.}\ \bibnamefont
  {Cuevas}}, \bibinfo {author} {\bibfnamefont {C.~P.}\ \bibnamefont {Lutz}},
  \bibinfo {author} {\bibfnamefont {C.~F.}\ \bibnamefont {Hirjibehedin}},\ and\
  \bibinfo {author} {\bibfnamefont {A.~J.}\ \bibnamefont {Heinrich}},\
  }\bibfield  {title} {\bibinfo {title} {Subgap structure in asymmetric
  superconducting tunnel junctions},\ }\href@noop {} {\bibfield  {journal}
  {\bibinfo  {journal} {Physical Review B}\ }\textbf {\bibinfo {volume} {74}},\
  \bibinfo {pages} {132501} (\bibinfo {year} {2006})}\BibitemShut {NoStop}%
\bibitem [{\citenamefont {Cuevas}\ \emph {et~al.}(1996)\citenamefont {Cuevas},
  \citenamefont {Mart{\'\i}n-Rodero},\ and\ \citenamefont
  {Yeyati}}]{cuevas1996hamiltonian}%
  \BibitemOpen
  \bibfield  {author} {\bibinfo {author} {\bibfnamefont {J.}~\bibnamefont
  {Cuevas}}, \bibinfo {author} {\bibfnamefont {A.}~\bibnamefont
  {Mart{\'\i}n-Rodero}},\ and\ \bibinfo {author} {\bibfnamefont {A.~L.}\
  \bibnamefont {Yeyati}},\ }\bibfield  {title} {\bibinfo {title} {Hamiltonian
  approach to the transport properties of superconducting quantum point
  contacts},\ }\href@noop {} {\bibfield  {journal} {\bibinfo  {journal}
  {Physical Review B}\ }\textbf {\bibinfo {volume} {54}},\ \bibinfo {pages}
  {7366} (\bibinfo {year} {1996})}\BibitemShut {NoStop}%
\bibitem [{\citenamefont {Zhu}\ \emph {et~al.}(2020)\citenamefont {Zhu},
  \citenamefont {Kong}, \citenamefont {Cao}, \citenamefont {Chen},
  \citenamefont {Papaj}, \citenamefont {Du}, \citenamefont {Xing},
  \citenamefont {Liu}, \citenamefont {Wang}, \citenamefont {Shen} \emph
  {et~al.}}]{zhu2020nearly}%
  \BibitemOpen
  \bibfield  {author} {\bibinfo {author} {\bibfnamefont {S.}~\bibnamefont
  {Zhu}}, \bibinfo {author} {\bibfnamefont {L.}~\bibnamefont {Kong}}, \bibinfo
  {author} {\bibfnamefont {L.}~\bibnamefont {Cao}}, \bibinfo {author}
  {\bibfnamefont {H.}~\bibnamefont {Chen}}, \bibinfo {author} {\bibfnamefont
  {M.}~\bibnamefont {Papaj}}, \bibinfo {author} {\bibfnamefont
  {S.}~\bibnamefont {Du}}, \bibinfo {author} {\bibfnamefont {Y.}~\bibnamefont
  {Xing}}, \bibinfo {author} {\bibfnamefont {W.}~\bibnamefont {Liu}}, \bibinfo
  {author} {\bibfnamefont {D.}~\bibnamefont {Wang}}, \bibinfo {author}
  {\bibfnamefont {C.}~\bibnamefont {Shen}}, \emph {et~al.},\ }\bibfield
  {title} {\bibinfo {title} {Nearly quantized conductance plateau of vortex
  zero mode in an iron-based superconductor},\ }\href@noop {} {\bibfield
  {journal} {\bibinfo  {journal} {Science}\ }\textbf {\bibinfo {volume}
  {367}},\ \bibinfo {pages} {189} (\bibinfo {year} {2020})}\BibitemShut
  {NoStop}%
\bibitem [{\citenamefont {Ivanchenko}\ and\ \citenamefont
  {Zil'berman}(1969)}]{ivanchenko1969josephson}%
  \BibitemOpen
  \bibfield  {author} {\bibinfo {author} {\bibfnamefont {Y.~M.}\ \bibnamefont
  {Ivanchenko}}\ and\ \bibinfo {author} {\bibfnamefont {L.}~\bibnamefont
  {Zil'berman}},\ }\bibfield  {title} {\bibinfo {title} {The josephson effect
  in small tunnel contacts},\ }\href@noop {} {\bibfield  {journal} {\bibinfo
  {journal} {Soviet Journal of Experimental and Theoretical Physics}\ }\textbf
  {\bibinfo {volume} {28}},\ \bibinfo {pages} {1272} (\bibinfo {year}
  {1969})}\BibitemShut {NoStop}%
\bibitem [{\citenamefont {Takayama}(1971)}]{takayama1971superconducting}%
  \BibitemOpen
  \bibfield  {author} {\bibinfo {author} {\bibfnamefont {H.}~\bibnamefont
  {Takayama}},\ }\bibfield  {title} {\bibinfo {title} {Superconducting
  fluctuation effects on the sin junction current},\ }\href@noop {} {\bibfield
  {journal} {\bibinfo  {journal} {Progress of Theoretical Physics}\ }\textbf
  {\bibinfo {volume} {46}},\ \bibinfo {pages} {1} (\bibinfo {year}
  {1971})}\BibitemShut {NoStop}%
\bibitem [{\citenamefont {Cea}\ and\ \citenamefont
  {Benfatto}(2016)}]{cea2016signature}%
  \BibitemOpen
  \bibfield  {author} {\bibinfo {author} {\bibfnamefont {T.}~\bibnamefont
  {Cea}}\ and\ \bibinfo {author} {\bibfnamefont {L.}~\bibnamefont {Benfatto}},\
  }\bibfield  {title} {\bibinfo {title} {Signature of the leggett mode in the a
  1 g raman response: From mgb 2 to iron-based superconductors},\ }\href@noop
  {} {\bibfield  {journal} {\bibinfo  {journal} {Physical Review B}\ }\textbf
  {\bibinfo {volume} {94}},\ \bibinfo {pages} {064512} (\bibinfo {year}
  {2016})}\BibitemShut {NoStop}%
\bibitem [{\citenamefont {Shim}\ \emph {et~al.}(2007)\citenamefont {Shim},
  \citenamefont {Yoon}, \citenamefont {Moodera},\ and\ \citenamefont
  {Hong}}]{shim2007all}%
  \BibitemOpen
  \bibfield  {author} {\bibinfo {author} {\bibfnamefont {H.}~\bibnamefont
  {Shim}}, \bibinfo {author} {\bibfnamefont {K.}~\bibnamefont {Yoon}}, \bibinfo
  {author} {\bibfnamefont {J.~S.}\ \bibnamefont {Moodera}},\ and\ \bibinfo
  {author} {\bibfnamefont {J.~P.}\ \bibnamefont {Hong}},\ }\bibfield  {title}
  {\bibinfo {title} {All mg b 2 tunnel junctions with al 2 o 3 or mgo tunnel
  barriers},\ }\href@noop {} {\bibfield  {journal} {\bibinfo  {journal}
  {Applied physics letters}\ }\textbf {\bibinfo {volume} {90}},\ \bibinfo
  {pages} {212509} (\bibinfo {year} {2007})}\BibitemShut {NoStop}%
\bibitem [{\citenamefont {Wan}\ \emph {et~al.}(2022)\citenamefont {Wan},
  \citenamefont {Dreher}, \citenamefont {Mu{\~n}oz-Segovia}, \citenamefont
  {Harsh}, \citenamefont {Guo}, \citenamefont {Mart{\'\i}nez-Galera},
  \citenamefont {Guinea}, \citenamefont {de~Juan},\ and\ \citenamefont
  {Ugeda}}]{wan2022observation}%
  \BibitemOpen
  \bibfield  {author} {\bibinfo {author} {\bibfnamefont {W.}~\bibnamefont
  {Wan}}, \bibinfo {author} {\bibfnamefont {P.}~\bibnamefont {Dreher}},
  \bibinfo {author} {\bibfnamefont {D.}~\bibnamefont {Mu{\~n}oz-Segovia}},
  \bibinfo {author} {\bibfnamefont {R.}~\bibnamefont {Harsh}}, \bibinfo
  {author} {\bibfnamefont {H.}~\bibnamefont {Guo}}, \bibinfo {author}
  {\bibfnamefont {A.~J.}\ \bibnamefont {Mart{\'\i}nez-Galera}}, \bibinfo
  {author} {\bibfnamefont {F.}~\bibnamefont {Guinea}}, \bibinfo {author}
  {\bibfnamefont {F.}~\bibnamefont {de~Juan}},\ and\ \bibinfo {author}
  {\bibfnamefont {M.~M.}\ \bibnamefont {Ugeda}},\ }\bibfield  {title} {\bibinfo
  {title} {Observation of superconducting collective modes from competing
  pairing instabilities in single-layer nbse2},\ }\href@noop {} {\bibfield
  {journal} {\bibinfo  {journal} {Advanced Materials}\ }\textbf {\bibinfo
  {volume} {34}},\ \bibinfo {pages} {2206078} (\bibinfo {year}
  {2022})}\BibitemShut {NoStop}%
\bibitem [{\citenamefont {Grinenko}\ \emph {et~al.}(2020)\citenamefont
  {Grinenko}, \citenamefont {Sarkar}, \citenamefont {Kihou}, \citenamefont
  {Lee}, \citenamefont {Morozov}, \citenamefont {Aswartham}, \citenamefont
  {B{\"u}chner}, \citenamefont {Chekhonin}, \citenamefont {Skrotzki},
  \citenamefont {Nenkov} \emph {et~al.}}]{grinenko2020superconductivity}%
  \BibitemOpen
  \bibfield  {author} {\bibinfo {author} {\bibfnamefont {V.}~\bibnamefont
  {Grinenko}}, \bibinfo {author} {\bibfnamefont {R.}~\bibnamefont {Sarkar}},
  \bibinfo {author} {\bibfnamefont {K.}~\bibnamefont {Kihou}}, \bibinfo
  {author} {\bibfnamefont {C.}~\bibnamefont {Lee}}, \bibinfo {author}
  {\bibfnamefont {I.}~\bibnamefont {Morozov}}, \bibinfo {author} {\bibfnamefont
  {S.}~\bibnamefont {Aswartham}}, \bibinfo {author} {\bibfnamefont
  {B.}~\bibnamefont {B{\"u}chner}}, \bibinfo {author} {\bibfnamefont
  {P.}~\bibnamefont {Chekhonin}}, \bibinfo {author} {\bibfnamefont
  {W.}~\bibnamefont {Skrotzki}}, \bibinfo {author} {\bibfnamefont
  {K.}~\bibnamefont {Nenkov}}, \emph {et~al.},\ }\bibfield  {title} {\bibinfo
  {title} {Superconductivity with broken time-reversal symmetry inside a
  superconducting s-wave state},\ }\href@noop {} {\bibfield  {journal}
  {\bibinfo  {journal} {Nature Physics}\ }\textbf {\bibinfo {volume} {16}},\
  \bibinfo {pages} {789} (\bibinfo {year} {2020})}\BibitemShut {NoStop}%
\bibitem [{\citenamefont {Gu}\ \emph {et~al.}(2022)\citenamefont {Gu},
  \citenamefont {Carroll}, \citenamefont {Wang}, \citenamefont {Ran},
  \citenamefont {Broyles}, \citenamefont {Siddiquee}, \citenamefont {Butch},
  \citenamefont {Saha}, \citenamefont {Paglione}, \citenamefont {Davis} \emph
  {et~al.}}]{gu2022detection}%
  \BibitemOpen
  \bibfield  {author} {\bibinfo {author} {\bibfnamefont {Q.}~\bibnamefont
  {Gu}}, \bibinfo {author} {\bibfnamefont {J.~P.}\ \bibnamefont {Carroll}},
  \bibinfo {author} {\bibfnamefont {S.}~\bibnamefont {Wang}}, \bibinfo {author}
  {\bibfnamefont {S.}~\bibnamefont {Ran}}, \bibinfo {author} {\bibfnamefont
  {C.}~\bibnamefont {Broyles}}, \bibinfo {author} {\bibfnamefont
  {H.}~\bibnamefont {Siddiquee}}, \bibinfo {author} {\bibfnamefont {N.~P.}\
  \bibnamefont {Butch}}, \bibinfo {author} {\bibfnamefont {S.~R.}\ \bibnamefont
  {Saha}}, \bibinfo {author} {\bibfnamefont {J.}~\bibnamefont {Paglione}},
  \bibinfo {author} {\bibfnamefont {J.}~\bibnamefont {Davis}}, \emph {et~al.},\
  }\bibfield  {title} {\bibinfo {title} {Detection of a pair density wave state
  in $ute_2$},\ }\href@noop {} {\bibfield  {journal} {\bibinfo  {journal}
  {arXiv preprint arXiv:2209.10859}\ } (\bibinfo {year} {2022})}\BibitemShut
  {NoStop}%
\bibitem [{\citenamefont {Koren}\ and\ \citenamefont
  {Lee}(2016)}]{koren2016observation}%
  \BibitemOpen
  \bibfield  {author} {\bibinfo {author} {\bibfnamefont {G.}~\bibnamefont
  {Koren}}\ and\ \bibinfo {author} {\bibfnamefont {P.~A.}\ \bibnamefont
  {Lee}},\ }\bibfield  {title} {\bibinfo {title} {Observation of two distinct
  pairs fluctuation lifetimes and supercurrents in the pseudogap regime of
  cuprate junctions},\ }\href@noop {} {\bibfield  {journal} {\bibinfo
  {journal} {Physical Review B}\ }\textbf {\bibinfo {volume} {94}},\ \bibinfo
  {pages} {174515} (\bibinfo {year} {2016})}\BibitemShut {NoStop}%
\bibitem [{\citenamefont {Hamidian}\ \emph {et~al.}(2016)\citenamefont
  {Hamidian}, \citenamefont {Edkins}, \citenamefont {Joo}, \citenamefont
  {Kostin}, \citenamefont {Eisaki}, \citenamefont {Uchida}, \citenamefont
  {Lawler}, \citenamefont {Kim}, \citenamefont {Mackenzie}, \citenamefont
  {Fujita} \emph {et~al.}}]{hamidian2016detection}%
  \BibitemOpen
  \bibfield  {author} {\bibinfo {author} {\bibfnamefont {M.}~\bibnamefont
  {Hamidian}}, \bibinfo {author} {\bibfnamefont {S.}~\bibnamefont {Edkins}},
  \bibinfo {author} {\bibfnamefont {S.~H.}\ \bibnamefont {Joo}}, \bibinfo
  {author} {\bibfnamefont {A.}~\bibnamefont {Kostin}}, \bibinfo {author}
  {\bibfnamefont {H.}~\bibnamefont {Eisaki}}, \bibinfo {author} {\bibfnamefont
  {S.}~\bibnamefont {Uchida}}, \bibinfo {author} {\bibfnamefont
  {M.}~\bibnamefont {Lawler}}, \bibinfo {author} {\bibfnamefont {E.-A.}\
  \bibnamefont {Kim}}, \bibinfo {author} {\bibfnamefont {A.}~\bibnamefont
  {Mackenzie}}, \bibinfo {author} {\bibfnamefont {K.}~\bibnamefont {Fujita}},
  \emph {et~al.},\ }\bibfield  {title} {\bibinfo {title} {Detection of a
  cooper-pair density wave in bi2sr2cacu2o8+ x},\ }\href@noop {} {\bibfield
  {journal} {\bibinfo  {journal} {Nature}\ }\textbf {\bibinfo {volume} {532}},\
  \bibinfo {pages} {343} (\bibinfo {year} {2016})}\BibitemShut {NoStop}%
\bibitem [{\citenamefont {Barlas}\ and\ \citenamefont
  {Varma}(2013)}]{barlas2013amplitude}%
  \BibitemOpen
  \bibfield  {author} {\bibinfo {author} {\bibfnamefont {Y.}~\bibnamefont
  {Barlas}}\ and\ \bibinfo {author} {\bibfnamefont {C.}~\bibnamefont {Varma}},\
  }\bibfield  {title} {\bibinfo {title} {Amplitude or higgs modes in d-wave
  superconductors},\ }\href@noop {} {\bibfield  {journal} {\bibinfo  {journal}
  {Physical Review B}\ }\textbf {\bibinfo {volume} {87}},\ \bibinfo {pages}
  {054503} (\bibinfo {year} {2013})}\BibitemShut {NoStop}%
\bibitem [{\citenamefont {Park}\ \emph {et~al.}(2021)\citenamefont {Park},
  \citenamefont {Cao}, \citenamefont {Watanabe}, \citenamefont {Taniguchi},\
  and\ \citenamefont {Jarillo-Herrero}}]{park2021tunable}%
  \BibitemOpen
  \bibfield  {author} {\bibinfo {author} {\bibfnamefont {J.~M.}\ \bibnamefont
  {Park}}, \bibinfo {author} {\bibfnamefont {Y.}~\bibnamefont {Cao}}, \bibinfo
  {author} {\bibfnamefont {K.}~\bibnamefont {Watanabe}}, \bibinfo {author}
  {\bibfnamefont {T.}~\bibnamefont {Taniguchi}},\ and\ \bibinfo {author}
  {\bibfnamefont {P.}~\bibnamefont {Jarillo-Herrero}},\ }\bibfield  {title}
  {\bibinfo {title} {Tunable strongly coupled superconductivity in magic-angle
  twisted trilayer graphene},\ }\href@noop {} {\bibfield  {journal} {\bibinfo
  {journal} {Nature}\ }\textbf {\bibinfo {volume} {590}},\ \bibinfo {pages}
  {249} (\bibinfo {year} {2021})}\BibitemShut {NoStop}%
\bibitem [{\citenamefont {Kim}\ \emph {et~al.}(2018)\citenamefont {Kim},
  \citenamefont {Wang}, \citenamefont {Nakajima}, \citenamefont {Hu},
  \citenamefont {Ziemak}, \citenamefont {Syers}, \citenamefont {Wang},
  \citenamefont {Hodovanets}, \citenamefont {Denlinger}, \citenamefont {Brydon}
  \emph {et~al.}}]{kim2018beyond}%
  \BibitemOpen
  \bibfield  {author} {\bibinfo {author} {\bibfnamefont {H.}~\bibnamefont
  {Kim}}, \bibinfo {author} {\bibfnamefont {K.}~\bibnamefont {Wang}}, \bibinfo
  {author} {\bibfnamefont {Y.}~\bibnamefont {Nakajima}}, \bibinfo {author}
  {\bibfnamefont {R.}~\bibnamefont {Hu}}, \bibinfo {author} {\bibfnamefont
  {S.}~\bibnamefont {Ziemak}}, \bibinfo {author} {\bibfnamefont
  {P.}~\bibnamefont {Syers}}, \bibinfo {author} {\bibfnamefont
  {L.}~\bibnamefont {Wang}}, \bibinfo {author} {\bibfnamefont {H.}~\bibnamefont
  {Hodovanets}}, \bibinfo {author} {\bibfnamefont {J.~D.}\ \bibnamefont
  {Denlinger}}, \bibinfo {author} {\bibfnamefont {P.~M.}\ \bibnamefont
  {Brydon}}, \emph {et~al.},\ }\bibfield  {title} {\bibinfo {title} {Beyond
  triplet: Unconventional superconductivity in a spin-3/2 topological
  semimetal},\ }\href@noop {} {\bibfield  {journal} {\bibinfo  {journal}
  {Science advances}\ }\textbf {\bibinfo {volume} {4}},\ \bibinfo {pages}
  {eaao4513} (\bibinfo {year} {2018})}\BibitemShut {NoStop}%
\bibitem [{\citenamefont {Tsuzuki}(1969)}]{tsuzuki1969effect}%
  \BibitemOpen
  \bibfield  {author} {\bibinfo {author} {\bibfnamefont {T.}~\bibnamefont
  {Tsuzuki}},\ }\bibfield  {title} {\bibinfo {title} {Effect of thermodynamic
  fluctuation of the superconductivity order parameter on the tunneling
  current},\ }\href@noop {} {\bibfield  {journal} {\bibinfo  {journal}
  {Progress of Theoretical Physics}\ }\textbf {\bibinfo {volume} {41}},\
  \bibinfo {pages} {1600} (\bibinfo {year} {1969})}\BibitemShut {NoStop}%
\end{thebibliography}%

\newpage

\appendix
\begin{widetext}
\section{Derivation of tunneling current due to collective modes  
}
We model the junction by the single-electron tunneling Hamiltonian
\be
H_{\textrm{tun}}=\sum_{\vec{k},\vec{p},\sigma}  t_{\vec{k},\vec{p}} e^{ieVt} c_{L,\vec{k},\sigma}^\dagger c^{}_{R,\vec{p},\sigma} + \textrm{h.c.},
\ee
where we have included the voltage $V$ across the junction through time-dependent tunneling. In this framework, one may assume the electrons of both sides to be in equilibrium and standard Matsubara diagrammatic techniques may be applied. The DC current is given by averaging the current operator related to $H_{\textrm{tun}}$ over time,
\be \label{currentmat}
I=\frac{1}{\beta}\int_0^\infty d\tau I(\tau) 
=-\frac{2e}{\beta}\sum_{kp\sigma} \textrm{Im} \int_0^\infty d\tau \langle T_\tau t_{\vec{k},\vec{p}} e^{i(ieV)\tau}c_{Lk\sigma}^\dagger(\tau) c_{Rp\sigma}(\tau) U(\beta)\rangle,
\ee
where we defined the imaginary time evolution operator
\be
U(\beta)=T_\tau \exp{[-\int_0^\beta d\tau' H_{\textrm{tun}}(\tau')]}.
\ee
(See Appendix B for a more careful treatment of the time-dependent factor in imaginary time.) We can now  expand Eq.\,\eqref{currentmat} to fourth order in $t_{\vec{k},\vec{p}}$ and proceed with the usual decoupling into products of Green functions. We are interested in diagrams that involve the imaginary time ordered pair propagator for the $R$ SC,
\be
\chi_{\alpha,\beta}(\vec{q},\tau)=-\langle T_\tau[\hat{\Delta}_{\vec{k}}^{(\alpha)}
(\vec{q},\tau)  \ \hat{\Delta}_\vec{k}^{(\beta) }(\vec{q},0)^\dagger]\rangle.
\ee
Upon Fourier transform and analytic continuation, $\chi_{\alpha,\beta}(\vec{q},i\omega_n\rightarrow\omega+i\eta)$, $\eta$ positive infinitesimal, becomes the retarded pair susceptibility defined in Eq.\,\eqref{pair2}. The relevant diagrams are shown in Fig.\,\ref{fig2}(c). They consist of the propagator connected to a triangular graph on each side. The sides of the triangle that connect to $\chi$ must refer to the $R$ SC, which means the remaining side refers to the $L$ SC. Furthermore, this side must couple to the anomalous propagator $F_L$ because we want to inject a pair from $L$ to $R$. There are two versions of the triangles, depending on whether the $L$ Green function is the normal $G$ or the anomalous $F$. We call the ones on the right of the pair propagator $M_1$ and $M_2$. It is easy to see that the ones on the left are the respective complex conjugates. In this way, we arrive at Eq.\,\eqref{M} where $M=M_1+M_2$. After analytic continuation, we obtain Eq.\,\eqref{currentstm}. Note that the voltage $eV$ is injected at each tunneling vertex and is passed directly to the pair propagator. This has the important consequence that the voltage gives a direct measure of the collective mode frequency via $2eV= \omega_i$.

\subsection{Collective mode: STM junction}

The momentum label on the triangles depends on the nature of the tunnel junction. We first discuss the STM or point contact tunneling case. The corresponding diagrams are shown in Fig.\,\ref{fig2}(a) and (b). To simplify notation, we consider a single s-wave component and suppress the $\alpha$  label. The tunneling occurs at a single point, which means that $t_{\vec{k},\vec{p}}$ is independent of momentum and can be set to a constant $\tilde{t}$. In contrast to  frequency, momentum is not conserved at the vertex for tunneling at a point. The integration over $k$ for the $L$ SC can be done separately, which simply gives the anomalous Green function at $\vec{r}=0$, i.e. $F_L(\vec{r}=0,\omega_n)= \pi N_L(0) \Delta_L/\sqrt{\omega_m^2+\Delta_L^2}$ where $N(0)$ is the density of states at the Fermi level. Furthermore, a finite momentum $q$ is passed on to the propagator, which needs to be integrated over. This gives rise to the integral over $\vec q$ shown in Eq.\,\eqref{currentJ}. This is the main difference between STM and planar tunneling, as shall be discussed below.

The full expression for $M(\vec{q},V)$ requires numerical integration. Here we consider some simple limits.  First we set $\vec q=0$ and then consider the expansion to quadratic order in $\vec q$.  Eq.\,\eqref{M} becomes
\begin{align}\label{Mapp}
M(\vec{q}=0,V=0)=&\ T\sum_m \Tilde{t}^2 F_L(\vec{r}=0,\omega_m)\int d\vec{p}\,[ G_R(\omega_m,\vec{p}) G_R(-\omega_m,-\vec{p})- F_R(\omega_m,\vec{p}) F_R(-\omega_m,-\vec{p})]   \\
=&\ T\sum_m \Tilde{t}^2\frac{\Delta_L}{\sqrt{\omega_m^2+\Delta_L^2}}N_L(0)N_R(0)\int d\xi_\vec{p} \left[\left( \frac{i\omega_m + \xi_\vec{p}}{\omega_m^2+E_\vec{p}^2}\right)\left(\frac{-i\omega_m + \xi_\vec{p}}{\omega_m^2+E_\vec{p}^2}\right) - \frac{\Delta_R^2}{(\omega_m^2+E_\vec{p}^2)^2}\right]    
\end{align}  
where $E_\vec{p}=\sqrt{\Delta^2+\xi_\vec{p}^2}$ and $\xi_\vec{p}=\epsilon(\vec{p})-\mu$. We have set $V$=0 in this expression because its dependence on $V$ is expected to be smooth for the following reason. As seen from the diagrams in Fig.\,\ref{fig2}, the effect of $V$ is to shift $\omega_m$ to $\omega_m+ieV$ and $\omega_m-ieV$ in the two normal Green functions $G$. The sum over $\omega_m$ in Eq.\,\eqref{Mapp} can be converted to an integral over the real variable $\omega_m \rightarrow \omega$ in the $T \rightarrow 0$ limit. It is clear that adding an imaginary part will affect the integral in a smooth way. 

The factor in square brackets in Eq.\,\eqref{Mapp} can be written as
\be
\frac{1}{\omega_m^2 +\xi_\vec{p}^2+\Delta_R^2}-\frac{2\Delta_R}{(\omega_m^2 +\xi_\vec{p}^2+\Delta_R^2)^2}=\frac{d}{d\Delta_R}\left[\frac{\Delta_R}{\omega_m^2 +\xi_\vec{p}^2+\Delta_R^2}\right].
\ee
On the other hand, the Josephson energy is given by \cite{ambegaokar1963tunneling}
\be
E_J=T \sum_m \Tilde{t}^2\int d\vec{k}d\vec{p}\, F_L(\omega_m,\vec{k})F_R(\omega_m,\vec{p})=
T \sum_m \Tilde{t}^2 N_L(0) N_R(0) \int  d\xi_\vec{p} \frac{\Delta_L}{\sqrt{\omega_m^2 +\Delta_L^2}}\frac{\Delta_R}{\omega_m^2 +\xi_\vec{p}^2+\Delta_R^2}.
\ee
Hence, we find that 
\be
M(\vec{q}=0,V=0) = \frac{\partial E_J}{\partial \Delta_R}
\ee
as quoted in the text.

Next we calculate the leading $q^2$ term in $M(\vec{q},V)$. We expand $\xi_{\vec{p}+\vec{q}} \approx \xi_\vec{p} + v_Fq \cos{\theta}$ where $\theta$ is the angle between $\vec{p}$ and $\vec{q}$. We keep only the linear in $q$ term in this expansion. This amounts to assuming a constant Fermi velocity $v_F$. After a straightforward but laborious calculation, we arrive at 
\begin{align}\label{Mq} 
M(\vec{q},V=0)-M(\vec{q}=0,V=0)\approx&\ -v_F^2q^2 \frac{\pi}{4}\Tilde{t}^2N_L(0)N_R(0)\int \frac{d\omega}{2\pi}\frac{\Delta_L}{\sqrt{\omega^2+\Delta_L^2}}\frac{\Delta_R^2-\omega^2}{(\Delta_R^2+\omega^2)^{5/2}} \\
=&\ -q^2\xi_R^2 \frac{\pi^2}{8}\Tilde{t}^2N_L(0)N_R(0)f(y=\Delta_L/\Delta_R) 
\end{align}  
where we defined the BCS coherence length of the $R$ SC $\xi_R=v_F/\pi \Delta_R$, as well as the dimensionless function 
\be  \label{fy}
f(y)=y\int dw \frac{1}{\sqrt{w^2+y^2}}\frac{1-w^2}{(1+w^2)^{5/2}}.
\ee
For $y \gg 1$, $f \rightarrow 2/3$ while for $y\ll 1$ it behaves as $f \sim y\ln{1/y}$. At $\Delta_L = \Delta_R$, $f(1) = \pi/4$. Thus, for $\Delta_L \gg \Delta_R$ and $\Delta_L \sim \Delta_R$, $M(\vec{q},V=0)$ decreases with $q$ on the scale of the inverse of the coherence length $\xi_R$. The physical picture is that the injected quasi-particle pair can travel a distance $\xi_R$ in the $R$ SC before forming a Cooper pair. In the $\Delta_L \ll \Delta_R$ limit the scale depends on $\Delta_L$ as well.

\begin{figure}
	\centering
	\includegraphics[width=0.55\textwidth]{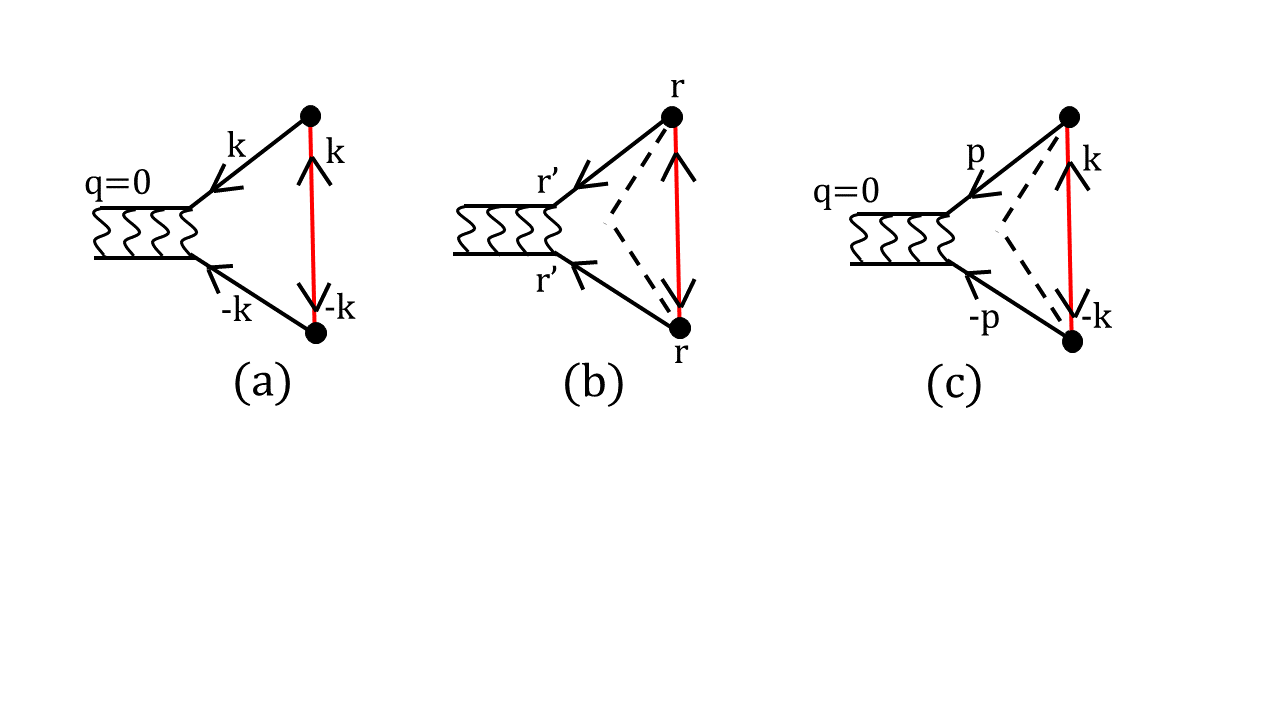}
	\caption{ Diagrams that contribute to $M_1$ in planar tunnel junctions. (a) momentum parallel to the plane is conserved. (b,c) diffusive case. Dashed line represents averaging over the random tunneling matrix elements. (b) is in position space. Note the the averaging forces both ends of the red line to be at the same $\vec{r}$. (c) The momentum space version of (b) obtained by Fourier transform.}
	\label{fig3}
	\vspace{-2mm}
\end{figure}

\subsection{Collective mode: Planar junction}

We now turn to the case of a planar tunnel junction. We consider two tunneling models. First, we assume that momentum parallel to the plane is conserved during the tunneling process. The momentum labels in the triangle graph are shown in Fig.\,\ref{fig3}(a). Note the difference to the STM case shown in Fig.\,\ref{fig2}. Now there is only one momentum integration. Importantly, the pair propagator carries momentum $q=0$. This will be changed to $q_H$ in the presence of a parallel magnetic field \cite{scalapino1970pair}. In contrast to the STM case, there is no convolution over $\vec{q}$. We thus recover the prediction from linear response to the pair tunneling Hamiltonian Eq.\,\eqref{pairH}.
Second, we consider diffusive scattering at the junction interface. We follow Takayama \cite{takayama1971superconducting} and assume local tunneling of the form
\be
H_{\textrm{local tun}}=\int d\vec{r }\,\hat{t}(\vec{r}) \sum_\sigma \psi_{L,\sigma}^\dagger (\vec{r}) \psi_{R,\sigma}^{}(\vec{r}) + \textrm{h.c.},
\ee
where $\vec{r}$ is the spatial coordinate in the plane, and $\hat{t}$ is a random variable such that $\langle\hat{t}(\Vec{r})\hat{t} (\Vec{r'})\rangle=t_0^2 \delta (\vec{r}-\vec{r'})$.  
This model has the desirable feature that the current, when computed to fourth order in $t_0$, scales correctly with the junction area. A more realistic model would have a nonzero average $\langle\hat{t}(\Vec{r})\rangle$. This latter part  conserves parallel momentum and thus may be included by simply adding the diagrams of Fig.\,\ref{fig3}(a) and (c). In Fig.\,\ref{fig3}(b) we show the triangle diagram in real space where the dashed line represents averaging over the random variable. Upon Fourier transform we obtain the diagram in momentum space as shown in Fig.\,\ref{fig3}(c). The momentum transfer to the pair propagator is zero in this case.  In a parallel magnetic field, momentum $\vec{q}_H$ is injected at the tunneling vertices, leading to $\vec{q}_H$ dependence of $M$ and excitation of the collective mode at $2\vec{q}_H$.
We conclude that in both cases, the planar tunneling case can be described by a pair tunneling Hamiltonian which is analogous to Eq.\,\eqref{pairH}, except that $C$ should be replaced by a slowly varying function of $V$ which is nonlocal in space. Indeed, this function is readily obtained from its Fourier transform $\Tilde{C}(\vec{q},V)=M(\vec{q},V)$, as described in the main text below Eq.\,\eqref{pairH2}.

\section{Calculation of leading MAR subgap structures 
}

In this section we employ the Matsubara method to treat the leading MAR structure which is fourth order in the tunneling matrix element. The equivalent calculation was done by Cuevas et al. \cite{cuevas1996hamiltonian} for the STM case using Keldysh technique. We first present results for the planar tunneling case, including also finite temperature effects. We were not able to find this case treated in the literature, and we find the Matsubara method to be less laborious than the Keldysh method, even though some subtlety is involved.

As noted in the text, the voltage drop across the junction can be absorbed into a time dependent tunneling matrix element  $t_{\vec{k},\vec{p}} e^{ieVt}$. After this step the SC in the leads are at equilibrium so that conventional Matsubara method can be used. This was done in calculating the matrix element $M$ that couples to the collective mode in the last section.  The proper procedure was given by Tsuzuki \cite{tsuzuki1969effect} and used by Takayama \cite{takayama1971superconducting}. The idea is to introduce $\Omega_n=2\pi nT$, where $n$ is a positive integer, and replace $t_{\vec{k},\vec{p}} e^{ieVt}$ by $t_{\vec{k},\vec{p}} e^{i\Omega_n \tau}$, where $\tau=it$. In this way the evolution in $\tau$ is unitary. At the end of the calculation, after internal Matsubara frequencies have been summed, we replace 
$i\Omega_n$ by $eV$. In a slight departure from Tsuzuki, we keep the adiabatic turning on term $e^{\eta t}$ as $e^{-i\eta \tau}$ so that $i\Omega_n \rightarrow eV+i\eta$. We will see that failure to follow this procedure will result in erroneous thermal factors. This subtlety did not arise in the calculation of the matrix element $M$ in the previous section which involves only  virtual excitations of quasi-particles. In contrast MAR involves the real excitation of quasi-particles and their thermal occupation appears in a crucial way.

\subsection{MAR: Planar junction}
To illustrate this point, we give some details for the case of planar junctions with diffusive scattering using the random tunneling model described in the last section \cite{takayama1971superconducting}. A representative diagram for the current that shows the flow of momentum and frequency is shown in \ref{fig4}(a).  
There are other diagrams where each solid line can be either $G$ or $F$ functions. In other words, the  diagram should be viewed as the trace of the matrix product of four Green function in Nambu space. In addition, the dashed line representing averaging over random tunneling matrix elements can connect the ends of the  solid lines on the top and bottom  instead of left and right as shown in Fig.\ref{fig4}a. However, its contribution can be shown to be less singular. Here, we show the evaluation of this particular diagram which we label by $i$. The sum over $\vec{k}$ and $\vec{k'}$ converts the anomalous Green function $F_L$ to the local form, and we have a single sum over $\vec{p}$. We write the contribution of this diagram to the MAR current for the diffusive planar junction as 
\be 
I^{\textrm{MAR}}_{\textrm{planar},i}=4e|t_0|^4 \textrm{Im} \ J_{i}(i\Omega_n \rightarrow eV+i\eta),
\ee
where
\begin{align} \label{J2d}
J_{i}(i\Omega_n)=&\ 
T\sum_m | F_L(x=0,\omega_m)|^2 A \int d\vec{p}\, [ G_R(\omega_m+\Omega_n,\vec{p}) G_R(-\omega_m+\Omega_n,-\vec{p}) \\
=&\ T\sum_m \frac{\pi^2 \Delta_L^2}{\omega_m^2+\Delta_L^2}  N^2_L(0) N_R(0) A \int d\xi_\vec{p}\, \left[ \frac{i\omega_m +i\Omega_n + \xi_\vec{p}}{(i\omega_m+i\Omega_n)^2 -E_\vec{p}^2} \right]\left[\frac{-i\omega_m +i\Omega_n + \xi_\vec{p}}{(-i\omega_m+i\Omega_n)^2 -E_\vec{p}^2} \right].
\end{align}
Here $A$ is the junction area. The sum over $m$ is done in the standard way by extending $i\omega_m$ to the complex $z$ and converting $\int d\xi$ to $\int dE \ E/\sqrt{E^2-\Delta^2}$ where
$E=\sqrt{\xi^2+\Delta^2}$. There are 6 poles $z_i$ in Eq.\,\eqref{J2d}.  The sum over $n$ leads to a sum over the product of the residues of these poles and $n_F(z_i)$ where $n_F$ is the Fermi function. For $eV > 0$ the important poles are at $z_3=i\Omega_n-E$ and $z_2=-z_3$.
The Fermi function becomes $n_F(z_3)=n_F(i\Omega_n-E)=n_F(-E)$.  The last step is crucial. Had we not replaced $eV$ by $i\Omega_n$, we would have gotten $n_F(eV-E)$ which is the wrong distribution and will give a thermal smearing for $eV$ near the gap $\Delta$.

After this step we can set $i\Omega_n  \rightarrow eV-i\eta$. A pole appears at $E=eV+i\eta$. Taking the imaginary part of  $J_{i}(i\Omega_n \rightarrow eV+i\eta)$ gives a delta function $\delta(2eV-2E)$. This is the expression of energy conservation: the tunneling of a Cooper pair gains an energy $2eV$ which are used to excite two quasi-particles each at energy $E$. This is because momentum is conserved on average after averaging over the random tunneling amplitudes and a pair of quasi-particles with opposite momenta $\vec{p}$ and $-\vec{p}$ 
 and equal energy $E$ are created. For $eV>\Delta$ this opens up a new threshold for conduction. The contribution of this diagram is given by 
\be  \label{I2d}
I^{\textrm{MAR}}_{\textrm{planar},i}=e\frac{\pi^3}{2}|t_0|^4 N^2_L(0)N_R(0) A \frac{eV}{\sqrt{(eV)^2-\Delta^2}}[n_F(-eV)-n_F(eV)]. 
\ee
Inclusion of the other diagrams only changes the numerical prefactor. The important thing to note is that the current appears as an inverse square root divergence above the threshold. This just reflects the density of states of finding the quasi-particle and is very different from the step singularity in the well known STM case. As we shall see below, this is because in that case a pair of quasi-particles at energy $E_p$ and $E_{p'}$ are excited and a convolution over two quasi-particle density of states appear. Also note that
 the thermal factor $n_F(-eV)-n_F(eV)$ comes from the poles at $z_3$ and $z_2$. This factor equals unity up to corrections of order $e^{-\Delta/T}$ which is negligible. Therefore there is no thermal smearing of the square root divergence near threshold.
 \begin{figure}
	\centering
	\includegraphics[width=0.55\textwidth]{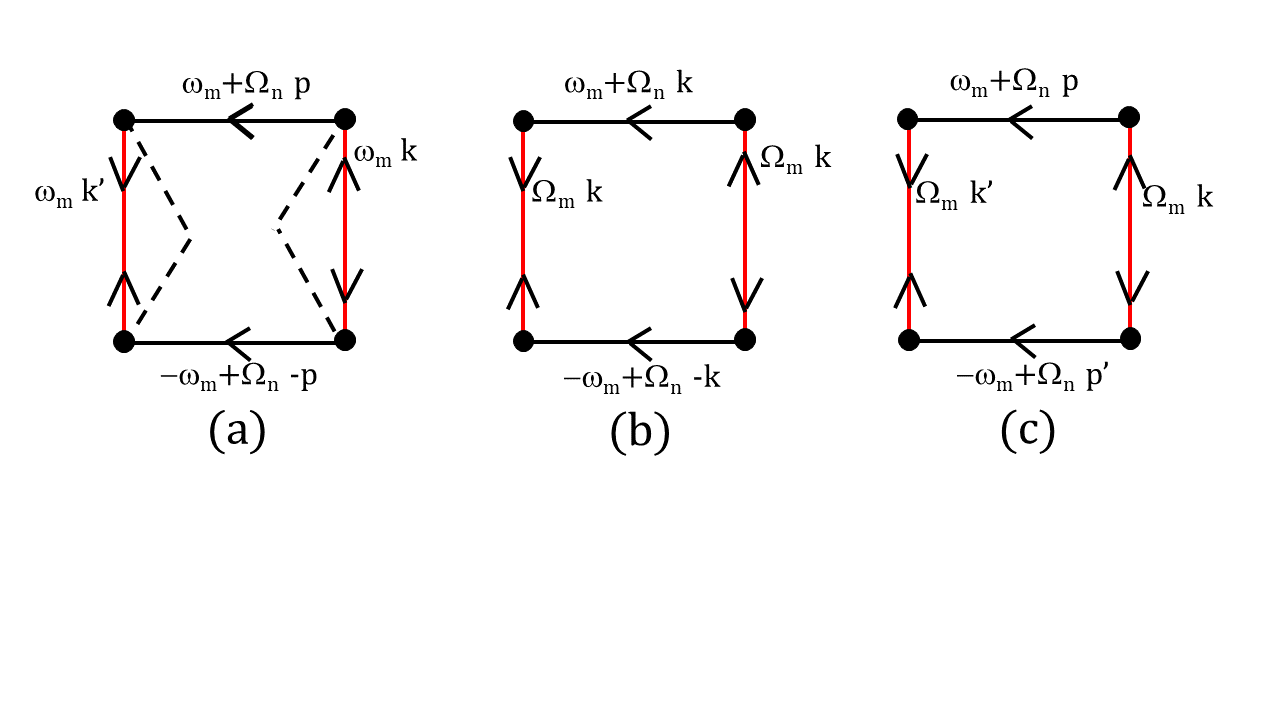}
	\caption{ Representative diagrams that contribute to the MAR currents. (a) Planar junction, diffusive case. Dashed line represents averaging over the random tunneling matrix elements. (b) Planar junction, clean case. Momentum parallel to the plane is conserved resulting in a single momentum integral over $k$. (c) STM case. The tunneling matrix element represented by the solid dots add or remove Matsubara frequency $\Omega_n$ which is analytically continued to $\Omega_n \rightarrow eV+i\eta$ after the sum over $\omega_m$ is performed. In addition to the diagrams shown, each solid line can be either $G$ or $F$ functions.}
	\label{fig4}
	\vspace{-2mm}
\end{figure}

 It is useful to express the results in terms of the conductance $1/R_N$  of the tunnel junction, which, in the diffusive case, is given by
 \be
 \frac{1}{R_N}= 4\pi \frac{e^2}{\hbar} |t_0|^2 N_L(0) N_R(0)A,
 \ee
 and consider the ratio of the MAR current to the normal state current at $eV=2
 \Delta$, $I_N=\frac{2\Delta}{eR_N}$.
 We also set $N_R(0)=\frac{m}{2\pi}=\frac{k_F^2}{4 \pi E_F}$ where $k_F$ and $E_F$ are the Fermi momentum and Fermi energy of the $R$ SC. We find
 \be  \label{I2dratio}
 \frac{I^{\textrm{MAR}}_{\textrm{planar}}}{I_N} \propto \frac{g}{Ak_F^2}\frac{E_F}{\Delta}\frac{eV}{\sqrt{(eV)^2-\Delta^2}}.
 \ee
 We can interpret $A k_F$ as the number of tunneling channels and $\mathcal{T}_{\textrm{eff}}=\frac{g}{Ak_F^2}$ as the tunneling probability per channel. $\mathcal{T}_{\textrm{eff}}$ describes the intrinsic transparency of a planar junction.

 We can repeat the calculation for the smooth planar junction where momentum parallel to the interface is conserved. A representative  diagram is shown in Fig.\,\ref{fig4}(b). Now there is only a single momentum sum and the results when expressed in terms of the conductance is essentially the same as Eq.\,\eqref{I2dratio}.

\subsection{MAR: STM junction}
We have also reproduced the result for the STM case. A representative diagram is shown in Fig.\,\ref{fig4}(c). Now there are two momentum sums over $\vec{p}$ and $\vec{p'}$, resulting in two energy integrals. Taking the imaginary part results in a delta function $\delta(E+E'=2eV)$. This expresses the fact that the energy gained by tunneling a pair is used to excite a pair of quasi-particles with momenta $\vec{p}$ and $\vec{p'}$ and energy $E$ and $E'$. Summing over all diagrams and taking $\Delta_L=\Delta_R=\Delta$, we find for $eV>\Delta$
\be \label{ISTM}
 I_{\textrm{STM}}^\textrm{MAR}= 2e\pi^3 |\Tilde{t}|^4 N^2_L(0) N^2_R(0)\Delta^2 \int_{\Delta} ^{2eV-\Delta}\ dE\ \frac{n_F(-E)-n_F(E)}{\sqrt{E^2-\Delta^2}\sqrt{(2eV-E)^2-\Delta^2}},
\ee
leading to a step function at $eV=\Delta$. This agrees with, and extends
Ref.\,\cite{cuevas1996hamiltonian} to finite temperature, demonstrating that the finite temperature correction is exponentially small in $\Delta/T$. Its ratio to the corresponding normal current $I_N$  is given by
\be
 \frac{I_{\textrm{STM}}^\textrm{MAR}}{I_N}=\frac{\pi}{16}g \ \theta(eV-\Delta)
\ee
Note that unlike the planar case, the factor $N_R(0)$ in the denominator, which led to the Fermi energy $E_F$ appearing in the numerator of Eq.\,\eqref{I2dratio}, does not arise.

We emphasize that there is no thermal smearing of the MAR peak until a temperature comparable to the energy gap $\Delta$ is reached. On the other hand, thermal noise gives rise to voltage fluctuations as discussed in the main text. The MAR structure can also be viewed as linear response to the order parameter in the $R$ SC except that quasi-particles are excited instead of a collective mode. Therefore the voltage fluctuation will broaden the MAR structure in the same way as the collective mode is broadened. Thus the minimum broadening is given by $\Gamma_0$. In practice, in planar junctions there are other sources of broadening such as local inhomogeneity. The latter is particularly important for many strongly correlated SCs such as cuprates and iron based SCs.
\end{widetext}

\end{document}